\documentclass[10pt, a4paper, copyright, logo, nonumbering]{instadeep}
\usepackage[listings]{tcolorbox}
\tcbuselibrary{listings, skins, breakable}
\usepackage{xcolor}

\definecolor{codebg}{RGB}{248,248,248}
\definecolor{codeframe}{RGB}{220,220,220}

\definecolor{pykeyword}{RGB}{0,0,150}    
\definecolor{pycomment}{RGB}{34,139,34}  
\definecolor{pystring}{RGB}{170,0,0}     
\definecolor{pynumber}{RGB}{128,0,128}   

\newtcblisting{codeblock}{
    listing only,
    breakable,
    colback=codebg,
    colframe=codeframe,
    arc=2pt,
    outer arc=2pt,
    boxrule=0.4pt,
    top=3pt,
    bottom=3pt,
    left=3pt,
    right=3pt,
    listing options={
        language=Python,
        basicstyle=\ttfamily\small,
        keywordstyle=\color{pykeyword}\bfseries,
        commentstyle=\color{pycomment}\itshape,   
        stringstyle=\color{pystring},
        numberstyle=\color{pynumber},
        showstringspaces=false,
        breaklines=true,
        columns=fullflexible,
    }
}
\title{MLIPAudit: A benchmarking tool for Machine Learned Interatomic Potentials}

\correspondingauthor{s.gutierrez@instadeep.com}

\keywords{Artificial Intelligence, Digital Biology, Systems Engineering}

\author[$\dagger$,1]{Leon Wehrhan}
\author[$\dagger$,1]{Lucien Walewski}
\author[1]{Marie Bluntzer}
\author[1]{Heloise Chomet}
\author[1]{Jules Tilly}
\author[1]{Christoph Brunken}
\author[*,1]{Silvia Acosta-Gutiérrez}

\affil[1]{InstaDeep}
\affil[$\dagger$]{Equal contribution}
\affil[*]{Corresponding author: s.gutierrez@instadeep.com}

\begin{abstract}
\textbf{Machine-learned interatomic potentials (MLIPs) promise to significantly advance atomistic simulations by delivering quantum-level accuracy for large molecular systems at a fraction of the computational cost of traditional electronic structure methods. While model hubs and categorisation efforts have emerged in recent years, it remains difficult to consistently discover, compare, and apply these models across diverse scenarios. The field still lacks a standardised and comprehensive framework for evaluating MLIP performance. We introduce MLIPAudit, an open, curated and modular benchmarking suite designed to assess the accuracy of MLIP models across a variety of application tasks. MLIPAudit offers a diverse collection of benchmark systems, including small organic compounds, molecular liquids, proteins and flexible peptides, along with pre-computed results for a range of pre-trained and published models. MLIPAudit also provides tools for users to evaluate their models using the same standardised pipeline. A continuously updated leaderboard tracks performance across benchmarks, enabling direct comparison on downstream tasks. By providing a unified, transparent reference framework for model validation and comparison, MLIPAudit aims to foster reproducibility, transparency, and community-driven progress in the development of MLIPs for complex molecular systems. In order to illustrate the use of the library, we present some benchmarks run on a series of internal models, along with publicly available ones (UMA-Small, MACE-OFF, MACE-MP). The library is available on \href{https://github.com/instadeepai/mlipaudit}{GitHub}, on \href{https://pypi.org/project/mlipaudit/}{PyPI} under the Apache license 2.0, and the \href{https://huggingface.co/spaces/InstaDeepAI/mlipaudit-leaderboard}{ leaderboard} can be accessed directly on HuggingFace}.
\end{abstract}

\begin{document}
\maketitle
\tableofcontents
\section{Introduction}
The accurate prediction of molecular and material properties is a cornerstone of scientific progress across disciplines, including drug discovery, functional material design, and process chemistry \cite{rupp2012fast,butler2018machine,ceder2013stuff}. Traditionally, this has been done using classical force fields, which enable efficient simulations of large systems relying on predefined functional forms and parameters derived from experiments or first-principles methods \cite{cornell1995second,mackerell1998all}. Although computationally inexpensive, classical force fields often struggle to capture complex chemical interactions or generalise beyond the systems for which they were parametrised. At the other end of the spectrum, first-principles methods such as density functional theory (DFT) offer higher accuracy but at significantly greater computational cost, typically limiting their use to systems with fewer than a few hundred atoms \cite{kohn1965self,parr1989density}. In recent years, machine-learned interatomic potentials (MLIPs) have emerged as a compelling middle ground. These models aim to retain the accuracy of first-principles methods while approaching the efficiency of classical force fields, by learning the potential energy surface directly from high-level electronic structure data \cite{lysogorskiy2021performant,batatia2022mace,kovacs2023evaluation,batzner2022e3,zaverkin2023transfer,haghighatlari2022newtonnet,shapeev2016moment,anstine2025aimnet2,kabylda2023efficient,behler2021four,musil2021physics,deringer2021gaussian,huang2021ab,daw1984embedded,deringer2021origins,baldwin2023dynamic,rosenbrock2021machine,mlip-library2025}.

Despite the rapid emergence of diverse MLIP architectures, which have significantly broadened the scope of atomistic simulations, the field continues to lack a standardised and rigorous framework for evaluating model performance in downstream applications. Many benchmarks focus on energy and force errors, which miss aspects like stability, transferability, and robustness. Recent works propose more holistic evaluations \cite{mlip_arena2024,fu2023forces,maxson2024reporting,ortner2022framework,waters2022benchmarking,zuo2019performance,kovacs2025maceoff, batzner2022e3, anstine2025aimnet2-rxn, brandon-uma2025}, which we detail in the Literature Review section. However, all these studies highlight the need for consistent and reproducible evaluation protocols that go beyond basic error metrics, aiming to establish benchmarking practices that reflect real-world simulation demands. Therefore, a universally adopted, comprehensive benchmarking suite that can guide both model development and deployment remains an open challenge for the community.

To address this gap, we introduce MLIPAudit: an open, curated repository of benchmarks, reference datasets, and model evaluations for MLIP models applied (in its first version) to the analysis of small molecules, molecular liquids and biomolecules. MLIPAudit is designed to complement model-centric testing by shifting the focus to systematic validation and comparison. It provides:

\begin{itemize}
    \item A diverse set of benchmark systems, including organic small molecules, flexible peptides, folded protein domains, molecular liquids and solvated systems.
    \item Pre-computed results for a range of published and pretrained MLIP models, enabling direct, fair comparisons.
    \item A continuously updated leaderboard, tracking performance across different tasks.
    \item A suite of tools for users to submit and evaluate their models within the same benchmarking pipeline. We support both Jax-based and Torch-based models, as long as they have an ASE \cite{ase-paper,ISI:000175131400009} calculator.

\end{itemize}

By providing a shared reference point for assessing accuracy, robustness, and generalisation, MLIPAudit aims to facilitate transparency, reproducibility, and community-wide progress in the development and deployment of MLIPs for complex molecular systems.

\section{Literature Review}
MLIP Audit aims to expand the existing methods and tools for benchmarking MLIPs. To put this work in context, we summarise current efforts for MLIP benchmarking here.

\textbf{Static regression metrics:} The first and most fundamental level of MLIP evaluation involves the use of standard regression metrics to quantify a model's ability to reproduce the reference quantum-mechanical (QM) data it was trained on. The most common benchmarks in this category are the root-mean-square-error (RMSE) and mean-absolute-error (MAE) calculated for energies and atomic forces on a held-out validation dataset \cite{morrow2023how}. These benchmarks are routinely reported with the release of new MLIP models, and state-of-the-art models achieve high accuracy on these tests. Although benchmarks for atomic energies and forces are a necessary baseline for the interpolation accuracy of the models, they are insufficient to estimate their practical utility. This is demonstrated, for example, by Gonzales et al. \cite{gonzales2024benchmarking}, who found that three models with very similar force validation error show significant variation in performance on a structural relaxation task.

\textbf{Assessment of physical and chemical behaviour}: Recent MLIP benchmarks generally accompany model releases and assess performance on physical and chemical properties using QM or experimental data, typically tailored to specific use cases. For models trained on small organic molecules, standard tests include dihedral scans, conformer selection, vibrational frequencies, and interaction energies \cite{christensen2021orbnet, kovacs2025maceoff, weber2025efficient}. Biomolecular benchmarks cover backbone sampling, water properties, and folding dynamics \cite{kovacs2025maceoff, weber2025efficient, wang2024enhancing}, while models trained on reactivity data are evaluated on their ability to reproduce product, reactant, and transition state geometries, as well as reaction pathways via string or NEB methods \cite{anstine2025aimnet2-rxn, levine_open_2025}.

Comparative studies have also emerged, evaluating multiple MLIPs across diverse benchmarks. Fu et al. \cite{fu2023forces} propose a suite spanning organic molecules, peptides, and materials, and find that models with low force errors may still perform poorly on simulation-based metrics like energy conservation and sampling. Similarly, Liu et al. \cite{liu2023discrepancies} report discrepancies in atom dynamics and rare events, even for models with strong regression accuracy. These findings reflect a growing consensus that static error metrics alone are insufficient for evaluating MLIPs, and that dynamic and simulation-based benchmarks are increasingly essential.

\textbf{Standardised benchmarks:} While a great variety of benchmarks for accurate physical and chemical properties can be collected from individual model releases and MLIP evaluation studies, a need remains for standardised benchmarks that can be used to compare models on a level playing field and get a holistic view of their utility regarding practical tasks.

This gap is addressed by leaderboards and standardised frameworks. MLIP Arena \cite{mlip_arena2024} is a leaderboard based on a benchmark platform focused on physical awareness, stability, reactivity, and predictive power. The framework comprises a small but well-selected suite of benchmarks that address known problems like data leakage, transferability, and overreliance on specific errors. Matbench Discovery \cite{riebesell2024matbench} features a leaderboard and evaluation framework that is easily extendable to additional models and focused exclusively on materials science. MOFSimBench \cite{kras_mofsimbench_2025} is a standardised benchmark specialised on metal-organic frameworks that highlights simulation metrics and bulk properties. MLIPX \cite{zills2025mlipx} provides a framework with a user-centric perspective, providing a set of reusable recipes that allow users to compose benchmarks for their specific tasks. 

These standardised frameworks are valuable tools to evaluate and compare MLIP models. However, they are limited to a specific domain of application, employ a small number of benchmarks or require development by the MLIP user. 

\section{Bridging the gap between model validation and downstream task}
As mentioned earlier, MLIPs have reached impressive levels of accuracy on carefully curated training datasets, yet their performance on real downstream tasks remains highly variable. A central difficulty lies in the fundamental mismatch between the data on which these models are trained and the regimes they are ultimately expected to simulate. Most training sets emphasize equilibrium or near-equilibrium configurations, where reference quantum data are most readily available and geometries remain close to chemically stable structures. In contrast, downstream applications, such as molecular dynamics (MD), kinetic modelling, or materials and molecular discovery, continuously drive systems into regions of configuration space that are sparsely represented or entirely absent from the training data. In these regimes, MLIPs must rely on extrapolation across poorly constrained areas of the potential energy surface (PES), making standard energy/force validation metrics only weakly predictive of the model’s ability to reproduce emergent physical behavior. As a result, downstream observables often correlate poorly with training loss, and models that look similar in static accuracy can diverge significantly during long-timescale simulations.

The lack of systematic, comparable benchmarking further complicates meaningful assessment.  Robustness to extrapolation, dynamical stability, and fidelity of long-timescale ensemble properties remain underexplored, despite being critical for the practical deployment of MLIPs in scientific workflows. The absence of consistent evaluation frameworks makes it difficult to identify genuine performance differences or diagnose model failure modes, hindering progress toward reliable, transferable potentials.

Recent efforts such as MACE-MP \cite{batatia2025foundationmodelatomisticmaterials} highlight both progress and remaining limitations. Built upon the MACE \cite{batatia2022mace} equivariant message-passing architecture, MACE-MP represents one of the most thoroughly benchmarked MLIP families to date across a broad inorganic materials space. The project provides transparent comparisons to both classical and machine-learned potentials, testing models on diverse crystalline structures, defect energetics, phonon spectra, and stability under MD. This represents a significant step toward reproducible and comprehensive benchmarking. However, MACE-MP remains focused primarily on crystalline inorganic systems. It offers limited coverage of molecules, liquids, surfaces, or heterogeneous systems—domains where sampling challenges, chemical diversity, and dynamical complexity are substantially greater. As such, MACE-MP does not yet constitute a general-purpose benchmark standard for MLIP evaluation across the wider materials and biochemical sciences.

A more complete benchmark framework, particularly for applications in biophysics and chemistry, requires evaluation criteria that reflect the complexity of biological and molecular environments. Such a benchmark should focused on biophysical validity capturing the molecular complexity and environmental diversity characteristic of biological systems. This includes representative conformations of proteins, nucleic acids, lipids, and small molecules, as well as heterogeneous environments such as water, ions, and membranes. It should also probe the model’s response to physically meaningful perturbations like rotamer changes and backbone motions, where subtle energetic differences drive functional outcomes. Beyond reproducing static structures, MLIPs must demonstrate long-timescale dynamical stability under physiologically relevant conditions, preserve accurate sampling behaviour, and generalise across diverse molecular families and assemblies. Standardized and reproducible metrics,including force and energy errors, MD drift, ensemble property fidelity, and stability tests, are essential for fair comparison. Practical considerations such as computational cost, scalability further determine whether a potential is usable in realistic biophysical workflows.

Constructing such a benchmark suite comes with significant caveats. Ensuring a truly blind test set is particularly challenging in biochemical systems, where structural similarity across sequences, folds, chemistries, and environments can easily lead to inadvertent data leakage. Achieving meaningful separation requires careful control of sequence identity, conformational diversity, and chemical coverage. Moreover, generating high-fidelity reference data at quantum-mechanical levels for large biomolecular systems is prohibitively expensive, and the long MD simulations required to evaluate stability and sampling substantially increase computational cost. Balancing scientific rigour with practical feasibility remains an open challenge.
\section{MLIPAudit Benchmarks}\label{benchmarks}
Bridging the gap between training accuracy and physical reliability requires benchmarking frameworks that evaluate MLIPs not only as data-driven regressors but also as physical models capable of supporting downstream simulation tasks. As previously discussed, benchmarking efforts already exist, particularly in the context of inorganic materials—comprehensive, cross-domain assessments remain limited. Our goal here is to contribute an initial step toward such broader evaluations, particularly for molecular and biochemical systems where benchmarking practices are still emerging.

\begin{figure}[!h]
    \centering
    \includegraphics[width=1.0\linewidth]{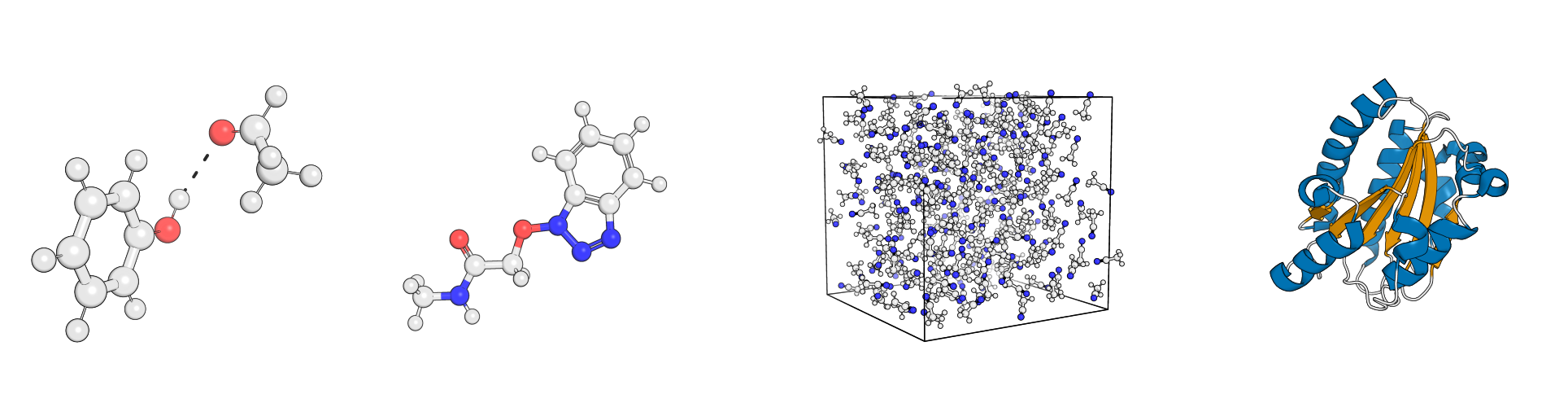}
    \caption{Representative molecular systems spanning increasing levels of structural and environmental complexity, from isolated dimers and drug-like molecules, to condensed-phase molecular liquids and folded biomolecules.}
    \label{fig:molecular-systems}
\end{figure}

To support a more rigorous and informative assessment of MLIP behaviour in these settings, MLIPAudit introduces a curated and modular suite of benchmarks spanning a range of molecular systems and complexity levels (Figure \ref{fig:molecular-systems}). The suite is designed to capture both general-purpose and domain-specific challenges relevant to real-world applications. Individual benchmark subsets highlight complementary aspects of model performance, including fundamental molecular dynamics stability, non-covalent interactions, conformational ranking of small organic molecules, and sampling of rotameric states in biomolecules. A detailed rationale for each benchmark category is provided below, covering: (i) general systems aimed at probing MD stability and scaling behaviour, (ii) small molecules relevant to materials chemistry, (iii) molecular liquids, and (iv) biomolecular systems. This first release provides a foundation upon which more comprehensive and increasingly standardized MLIP benchmarking frameworks can be built.

In the following subsections, we describe the composition, rationale, and evaluation criteria for each benchmark category.
\subsection{General benchmarks}
The general benchmarks implemented in MLIPAudit are system-agnostic and focus on fundamental molecular dynamics (MD) stability and performance metrics that are applicable across molecular systems. Two benchmarks are included in this category:
\begin{itemize}
    \item \textbf{Stability}: assesses the dynamical stability of an MLIP during an MD simulation for a diverse set of large biomolecular systems. For each system, the benchmark performs an MD simulation using the MLIP model in the NVT ensemble at 300 K for 100,000 steps (100~ps), leveraging the jax-md engine, as integrated via the mlip library\cite{mlip-library2025}. The test monitors the system for signs of instability by detecting abrupt temperature spikes (“explosions”) and hydrogen atom drift. These indicators help determine whether the MLIP maintains stable and physically consistent dynamics over extended simulation times.
    \vspace{4pt}
    \item \textbf{Inference Scaling}: evaluates how the computational cost of an MLIP scales with the system size. By running single, long MD episodes on a series of molecular systems of increasing size, we systematically assess the relationship between molecular complexity and inference performance. This benchmark is not used for scoring, but it aims at helping the user to pick the best model in terms of time-to-solution for the application task.
\end{itemize}
\subsection{Small Molecules}
MLIPAudit small-molecule benchmarks focus on the ability of MLIPs to reproduce the properties and dynamics of small organic molecules, including their conformational sampling and interactions with other molecules. In order of task complexity:
    
\begin{itemize}
    \item \textbf{Bond Length}: evaluates the ability of MLIPs to accurately model the equilibrium bond lengths of small organic molecules during MD simulations. This is an important test to understand whether the MLIP respects basic chemistry throughout simulations. Accurate prediction of bond length is crucial for capturing the structural and electronic properties of any chemically relevant compounds. For each molecule in the dataset, the benchmark performs an MD simulation with the same configuration described in the stability benchmark. Throughout the trajectory, the positions of the bond atoms are tracked, and their deviation from a reference bond length of the QM-optimised starting structure is calculated. The average deviation over the trajectory provides a direct measure of the MLIP’s ability to maintain bond lengths under thermal fluctuations, enabling quantitative comparison to reference data or other models.
    \vspace{4pt}
    \item \textbf{Ring Planarity}: evaluates the ability of MLIPs to preserve the planarity of aromatic and conjugated rings in small organic molecules during molecular dynamics simulations. Aromatic rings (e.g., benzene) are inherently planar due to delocalised $\pi$ electrons. Ring planarity enforcement is crucial in molecular dynamics simulations because it preserves the correct geometry, electronic structure, and interactions of aromatic and conjugated systems. Without proper planarity (e.g., via improper torsions), simulations can produce chemically unrealistic distortions that compromise accuracy in energy, flexibility, and binding predictions. This is especially important in molecules like benzene, tyrosine side chains, nucleobases, and drug scaffolds, where planarity governs stacking, hydrogen bonding, and overall stability. For each molecule in the dataset, the benchmark performs an MD simulation with the same configuration described in the stability benchmark. Throughout the trajectory, the positions of the ring atoms are tracked, and their deviation from a perfect plane is quantified using the root mean square deviation (RMSD) from planarity. The ideal plane of the ring is computed using a principal component analysis of the ring’s atoms. The average deviation over the trajectory provides a direct measure of the MLIP’s ability to maintain ring planarity under thermal fluctuations, enabling quantitative comparison to reference data or other models.
    \vspace{4pt}
    \item \textbf{Dihedral Scan}: evaluates the MLIP’s ability to reproduce torsional energy profiles of rotatable bonds in small molecules, aiming to approach the quantum-mechanical QM reference quality. Dihedral scans are essential for mapping how a molecule’s energy changes as bonds rotate, revealing preferred conformations and energy barriers. Beyond force field development, they are also used in studying reaction mechanisms, analysing conformational dynamics in drug discovery, validating quantum chemistry methods, and guiding the design of flexible or constrained molecules. For each molecule, the benchmark leverages the mlip library for model inference, comparing the predicted energies along a dihedral scan to QM reference energy profiles. The reference profile is shifted so that its global minimum is zero, and the MLIP profile is aligned to the same conformer. Performance is quantified using the following metrics: MAE and RMSE. The Pearson correlation coefficient between the MLIP-predicted and reference datapoints and the mean barrier height error.
    \vspace{4pt}
    \item \textbf{Non-covalent Interactions}: tests if the MLIP can reproduce interaction energies of molecular complexes driven by non-covalent interactions. Non-covalent interactions are of the highest importance for the structure and function of every biological molecule. This benchmark assesses a broad range of interaction types: London dispersion, hydrogen bonds, ionic hydrogen bonds, repulsive contacts and sigma hole interactions. Assessing the accuracy of non-covalent interactions is crucial for evaluating how well computational models capture key forces like hydrogen bonding, $\pi$-$\pi$ stacking, and van der Waals interactions that govern molecular recognition, binding, and assembly. This is essential not only for force field development, but also for validating quantum methods, guiding molecular design, modelling biomolecular interfaces, and studying condensed-phase behaviour such as solvation and aggregation. The benchmark runs energy inference on all structures of the distance scans of bi-molecular complexes in the dataset. The key metric is the RMSE of the interaction energy, which is the minimum of the energy well in the distance scan, relative to the energy of the dissociated complex,  compared to the reference data. For repulsive contacts, the maximum of the energy profile is used instead. Some of the molecular complexes in the benchmark dataset contain exotic elements (see dataset section). In case the MLIP has never seen an element of a molecular complex, this complex will be skipped in the benchmark.
    \vspace{4pt}
    \item \textbf{Reference Geometry Stability}: assesses the MLIP’s capability to preserve the ground-state geometry of organic small molecules during energy minimisation, ensuring that initial DFT-optimised structures remain accurate and physically consistent. Each system is minimised using the Broyden–Fletcher–Goldfarb–Shanno (BFGS) algorithm (ASE default parameters). After minimisation, structural fidelity is assessed by computing the RMSD of all heavy atoms relative to the initial geometry, using the RMSD implementation provided by mdtraj \cite{McGibbon2015MDTraj}.
    \vspace{4pt}
    \item \textbf{Conformer Selection}: evaluates the MLIP’s ability to identify the most stable conformers within an ensemble of flexible organic molecules and accurately predict their relative energy differences. It focuses on capturing subtle intramolecular interactions and strain effects that influence conformational energies. These metrics assess both numerical accuracy and the MLIP’s ability to preserve relative conformer energetics, which is crucial for downstream applications such as conformational sampling and compound ranking.
    \vspace{4pt}
    \item \textbf{Tautomers}: assesses the ability of MLIP to accurately predict the relative energies and stabilities of tautomeric forms of small molecules in vacuum. Tautomers are structural isomers that interconvert via proton transfer and/or double bond rearrangement, and accurately estimating the energy gap between them is an important measure of chemical accuracy in the MLIP framework. Tautomer ranking assesses a model’s ability to predict the relative stability of different tautomeric forms of a molecule, which is critical for accurately modelling protonation states, reactivity, and binding affinities. It is especially important in drug discovery, quantum method benchmarking, and cheminformatics, where tautomers can dramatically affect molecular properties and biological activity. For each molecule, the benchmark compares MLIP-predicted energies against QM reference data. Performance is quantified by comparing the absolute deviation of the energy difference between the tautomeric forms from the DFT data.
    \vspace{4pt}
    \item \textbf{Reactivity}: assesses the MLIP’s capability to model chemical reactivity. The reactivity-tst benchmark tests the ability to predict the energy of transition states relative to the reaction's reactants and products and thereby the activation energy and enthalpy of a reaction. This benchmark calculates the energy of reactants, products and transition states of a large dataset of reactions. From the difference between these states, the activation energy and enthalpy of formation can be calculated. The performance is quantified using the MAE and RMSE in activation energy and enthalpy of formation. The reactivity-neb benchmark evaluates the capability to converge a set of nudged elastic band calculations with a known transition state. The performance is quantified by the percentage of converged calculations.
\end{itemize}

\subsection{Molecular Liquids}
The MLIPAudit molecular liquids benchmark focuses on assessing long-range interactions by computing the radial distribution function for different molecular liquids.
\begin{itemize}
    \item \textbf{Radial Distribution Function}: assesses the ability of MLIP to accurately reproduce the radial distribution function (RDF) of liquids. The RDF characterises the local and intermediate-range structure of a liquid by describing how particle density varies as a function of distance from a reference particle. Accurate modelling of the RDF is essential for capturing both short-range ordering and long-range interactions, which are critical for understanding the microscopic structure and emergent properties of liquid systems. The benchmark performs an MD simulation using the MLIP model in the NVT ensemble at 300 K for 500,000 steps, leveraging the jax-md engine from the mlip library. The starting configuration is already equilibrated. For every specific atom pair (e.g., oxygen-oxygen in water), the radial distribution function (RDF or g(r)) is calculated from the simulation, as:
    \begin{equation}
        g(r)=\frac{1}{4\Pi r^{2} \rho N}\langle \sum_{i=1}^{N} \sum_{j\neq i}^{N} \delta (r-r_{ij})\rangle
    \end{equation}
    where: $r$ is the distance from a reference particle,$\rho$ is the average number density, $N$ is the number of particles, $r_{ij}$ is the distance between particles and $\delta$ is the Dirac delta function. Angle brackets denote an ensemble average.
    For each test case, the benchmark computes $r_{\text{peak}} = \underset{r}{\operatorname{arg\,max}}\, g(r)$ and compares it with the experimental value for the first solvation shell.

\end{itemize}

\subsection{Biomolecules}
MLIPAudit biomolecule benchmarks focus on assessing the properties and dynamics of proteins, including their folding behaviour, structural stability, and conformational sampling.

\begin{itemize}
    \item \textbf{Protein Folding Stability}:  evaluates the ability of an MLIP to preserve the structural integrity of experimentally determined protein conformations during MD simulations. It assesses the retention of secondary structure elements and overall compactness across a set of known protein structures. This module analyses the folding trajectories of proteins in MLIP simulations. For each molecule in the dataset, the benchmark performs an MD simulation with the same configuration described in the stability benchmark. We track how Root Mean Square Deviation (RMSD), TM Score \cite{TMscore}, Dictionary of Secondary Structure in Proteins (DSSP) \cite{dssp} and Radius of Gyration change over time.
    \vspace{4pt}
    \item \textbf{Sampling Outlier Detection}: Assesses the structural quality of sampled conformations by computing backbone Ramachandran angles ($\phi/\psi$) and side-chain rotamer angles ($\chi$), and identifying outliers through comparison with reference rotamer libraries \cite{lovell2000}. For each molecule in the dataset, the benchmark performs an MD simulation with the same configuration described in the stability benchmark. The outlier detection identifies residues whose dihedral angles fall outside expected ranges, relying on the fast KDtree \cite{kdtree} scipy \cite{2020SciPy-NMeth} implementation. The analysis provides a global percentage of outliers for backbone and rotamers per structure, as well as a more detailed analysis per residue type.
\end{itemize}

For each benchmark, a set of test cases has been curated (Appendix \ref{appendix:sup-fig-tab}, Table \ref{tab:test-cases}). It must be noted that as public datasets increase, it becomes increasingly challenging to ensure zero overlap between the training data and the relevant chemistry that one needs to include to ensure the relevance and reliability of the benchmarks. 
\section{Benchmarks Library Overview}
\subsection{Purpose and Design Philosophy}
The purpose of the \textit{mlipaudit} library is to provide users with a suite of benchmarks that help them decide which is the best MLIP model for their downstream application. The \textit{mlipaudit}
library was built in accordance with the following key design principles:
\begin{itemize}
    \item \textbf{Ease-of-use:} The library should be simple to install and use, especially for users that do not plan to add new benchmarks but instead run the existing benchmarks for their trained MLIP models.
    \item \textbf{Extensibility:} If required, \textit{mlipaudit} can be a extended with new benchmarks.
    \item \textbf{Visualisation:} The library is designed to include a UI for visualisation of results. It is easy-to-use for the existing benchmarks, and easy to extend for newly developed benchmarks.
\end{itemize}

\subsection{Using the CLI Tool}

After installation and activating the respective Python environment, the command line tool \texttt{mlipaudit} should be available with two tasks:
\begin{itemize}
    \item \texttt{mlipaudit benchmark}: The benchmarking CLI task. It runs the full or partial benchmark suite for one or more models. Results will be stored locally in multiple JSON files in an intuitive directory structure.
    \item \texttt{mlipaudit gui}: The UI app for visualisation of the results. Running it opens a browser window and displays the web app. Implementation is based on \href{https://streamlit.io/}{streamlit}.
\end{itemize}

This CLI tool is the easiest way to run all our benchmarks for your models without the requirement to write any additional Python code.
We recommend to use the \texttt{-h/-{}-help} option of the CLI to learn more about the available configurability of the command. We also refer to the tutorials in our code documentation for more details on the usage.

We provide three ways to interface MLIP models, (1) by specifying a path to a ".zip" file containing the model compatible with the \href{https://github.com/instadeepai/mlip}{mlip} library, or alternatively, by providing a path to a Python file that holds some simple code to instantiate either (2) a \texttt{ForceField} object from the \href{https://github.com/instadeepai/mlip}{mlip} library, or (3) a \texttt{Calculator} object from the \href{https://ase-lib.org/index.html}{ASE} library. Again, we refer to our code documentation for more details. Using option (3) even allows for running Torch-based models, however, we emphasise that in this case, the highly efficient JAX-MD integration of the \href{https://github.com/instadeepai/mlip}{mlip} library cannot be used and significantly lower simulation speed should be expected.

\subsection{Code Overview -- Basic Pattern for Running Benchmarks}
As can be seen in our main benchmarking script located at \texttt{src/mlipaudit/main.py} in our open-source repository (can be run via CLI tool), the basic pattern for running benchmarks with \textit{mlipaudit} is the following:

\begin{codeblock}
from mlipaudit.benchmarks import TautomersBenchmark
from mlipaudit.io import write_benchmark_result_to_disk
from mlip.models import Mace
from mlip.models.model_io import load_model_from_zip

force_field = load_model_from_zip(Mace, "./mace.zip")

benchmark = TautomersBenchmark(force_field)
benchmark.run_model()
result = benchmark.analyze()

write_benchmark_result_to_disk(
    TautomersBenchmark.name, result, "./results/mace"
)
\end{codeblock}

After initialising a benchmark class, in this example \texttt{TautomersBenchmark}, we call \verb|run_model()| to execute all inference calls and simulations required with the MLIP force field model. The raw output of this is stored inside the class instance. Next, we run \verb|analyze()| to produce the final benchmarking results. This function returns the results class, which is always a derived class of \texttt{BenchmarkResult}, in this example, \texttt{TautomersResult}. The function \verb|write_benchmark_result_to_disk| then writes these results to disk in a standardised JSON format.

\subsection{How to Contribute a Benchmark}
A new benchmark class can easily be implemented as a derived class of the abstract base class \texttt{Benchmark}. The attributes and members to override are:
\begin{itemize}
    \item \verb|name|: A unique name for the benchmark.
    \item \verb|category|: A string that represents the category of the benchmark. If not overridden, “General” is used. Currently, used exclusively for visualisation in the GUI.
    \item \verb|result_class|: A reference to the results class of the benchmark. More details below.
    \item \verb|model_output_class|: A reference to the model output class of the benchmark. More details below.
    \item \verb|required_elements|: A set of element symbols that are required by a model to run this benchmark.
    \item \verb|skip_if_elements_missing|: Boolean that has a default of \texttt{True} and hence does not need to be overridden. However, if you want your benchmark to still run even if a model is missing some required elements, then this should be overridden to be \texttt{False}. A reason for this would be that parts of the benchmark can still be run in this case and the missing elements will be handled on a case-by-case basis inside the benchmark’s run function.
    \item \verb|run_model|: This method implements running all inference calls and simulations related to the benchmark. This method can take a significant time to execute. As part of this, the raw output of the model should be stored in a model output class that needs to be implemented and must be derived from the base class \verb|ModelOutput|, which is a \href{https://docs.pydantic.dev/latest/}{pydantic} model (works similar to dataclasses but with type validation and serialisation built in). The model output of this type is then assigned to an instance attribute \verb|self.model_output|.
    \item \verb|analyze|: This method implements the analysis of the raw results and returns the benchmark results. This works similarly to the model output, where the results are a derived class of \verb|BenchmarkResult| (also a \href{https://docs.pydantic.dev/latest/}{pydantic} model).
\end{itemize}

Hence, to add a new benchmark, three classes must be implemented, the benchmark, model output, and results class.

Note that we also recommend to implement a very minimal version of the benchmark itself that is run if \verb|self.run_mode == RunMode.DEV|. For very long-running benchmarks, we also recommend to implement a version for \verb|self.run_mode == RunMode.FAST| that may differ from \verb|self.run_mode == RunMode.STANDARD|, however, for most benchmarks this may not be necessary.

Minimal example implementation
Here is an example of a very minimal new benchmark implementation:

\begin{codeblock}
import functools
from mlipaudit.benchmark import Benchmark, BenchmarkResult, ModelOutput

class NewResult(BenchmarkResult):
    errors: list[float]

class NewModelOutput(ModelOutput):
    energies: list[float]

class NewBenchmark(Benchmark):
    name = "new_benchmark"
    category = "New category"
    result_class = NewResult
    model_output_class = NewModelOutput
    required_elements = {"H", "N", "O", "C"}

    def run_model(self) -> None:
        energies = _compute_energies_blackbox(self.force_field, self._data)
        self.model_output = NewModelOutput(energies=energies)

    def analyze(self) -> NewResult:
        score, errors = _analyze_blackbox(self.model_output, self._data)
        return NewResult(score=score, errors=errors)

    @functools.cached_property
    def _data(self) -> dict:
        data_path = self.data_input_dir / self.name / "new_benchmark_data.json"
        return _load_data_blackbox(data_path)
\end{codeblock}

The data loading as a cached property is only recommended if the loaded data is needed in both the \verb|run_model()| and the \verb|analyze()| functions.

Note that the functions \verb|_compute_energies_blackbox| and \verb|_analyze_blackbox| are placeholders for the actual implementations.

Another class attribute that can be specified optionally is \verb|reusable_output_id|, which is \verb|None| by default. It can be used to signal that two benchmarks use the exact same \verb|run_model()| method and the exact same signature for the model output class. This ID should be of type tuple with the names of the benchmarks in it, see the benchmarks \verb|Sampling| and \verb|FoldingStability| for an example of this. See the source code of the main benchmarking script for how it reuses the model output of one for the other benchmark without rerunning any simulation or inference.

Furthermore, an import for the new benchmark in the \verb|src/mlipaudit/benchmarks/__init__.py| file needs to be added such that the benchmark can be automatically picked up by the CLI tool.

\textbf{Data}: The benchmark base class downloads the input data for a benchmark from \verb|HuggingFace| automatically if it does not yet exist locally. As you can see in the minimal example above, the benchmark expects the data to be in the directory \verb|self.data_input_dir / self.name|. Therefore, if you place your data in this directory before initialising the benchmark, it will not try to download anything from \verb|HuggingFace|. This mechanism allows the data to be provided in custom ways.

\textbf{UI page}: To create a new benchmark UI page, we refer to the existing implementations which are located in \verb|src/mlipaudit/ui| for how to add a new one. The basic idea is that a page is represented by a function like this:

\begin{codeblock}
def new_benchmark_page(
    data_func: Callable[[], dict[str, NewResult]],
) -> None:
    data = data_func()  # data is a dictionary of model names and results

    # add rest of UI page implementation here
    pass
\end{codeblock}

The implementation must be a valid \href{https://streamlit.io/}{streamlit} page.

In order for this page to be automatically included in the \verb|UI app|, you need to wrap this new benchmark page in a derived class of \verb|UIPageWrapper| like this,

\begin{codeblock}
class NewBenchmarkPageWrapper(UIPageWrapper):

    @classmethod
    def get_page_func(cls):
        return new_benchmark_page

    @classmethod
    def get_benchmark_class(cls):
        return NewBenchmark
\end{codeblock}

and then make sure to add the import of your new benchmark page to the \verb|src/mlipaudit/ui/__init__.py| file. This will result in your benchmark’s UI page being automatically picked up and displayed.
 
Note that as you need to modify some existing source code files of \textit{mlipaudit} to include your new benchmarks, this cannot be achieved purely with the pip installed library, however, we recommend to clone or fork our repository and run this local version instead after adding your own benchmarks with minimal code changes. See our code documentation for more information.

\section{Benchmarks Scoring}
\label{appendix-scoring}
To enable consistent and fair comparison across models, we define a composite score that aggregates performance over all compatible benchmarks. Each benchmark \( b \in \mathcal{B} \) may report one or more metrics \( x_{m,b}^{(i)} \), where \( i = 1, \ldots, N_b \) indexes the \( N_b \) metrics evaluated for the model \( m \). For each metric, we compute a normalised score using a soft thresholding function based on a DFT-derived reference tolerance \( t_b^{(i)} \) (see Table \ref{tab:benchmark-acceptable-ranges}):

\[
s_{m,b}^{(i)} =
\begin{cases}
1, & \text{if } x_{m,b}^{(i)} \leq t_b^{(i)} \\
\exp\left(-\alpha \cdot \frac{x_{m,b}^{(i)} - t_b^{(i)}}{t_b^{(i)}}\right), & \text{otherwise}
\end{cases}
\]

where \( \alpha \) is a tunable parameter controlling the steepness of the penalty (e.g., \( \alpha = 3 \)). The per-benchmark score is then computed as the average over all its metric scores:

\[
s_{m,b} = \frac{1}{N_b} \sum_{i=1}^{N_b} s_{m,b}^{(i)}
\]

Let \( \mathcal{B}_m \subseteq \mathcal{B} \) denote the subset of benchmarks for which the model \( m \) has valid data (i.e., benchmarks compatible with its element set). The final model score is the mean over all benchmarks on which the model could be evaluated:

\[
S_m = \frac{1}{|\mathcal{B}_m|} \sum_{b \in \mathcal{B}_m} s_{m,b}
\]

This scoring framework ensures that models are rewarded for meeting or exceeding DFT-level accuracy. In the current version, full benchmarks are skipped if a model does not have all the necessary chemical elements to run all the test cases. This is true for all benchmarks, but non-covalent interactions, in which we do a per-test-case exception. Benchmarks with multiple metrics contribute proportionally, and the result is a single interpretable score \( S_m \in [0,1] \) that balances physical fidelity, chemical coverage, and overall model robustness. The thresholds for the different benchmarks have been chosen based on the literature. In the case of tautomers, energy differences are very small; therefore, we've chosen a stricter threshold of 1-2 kcal/mol, which is not enough for classification. Thresholds for biomolecules are borrowed from traditional literature in molecular modelling.

\begin{table}[!h]
\centering
\small
\caption{Score thresholds across benchmarks.}
\begin{tabular}{l|p{4.5cm}|p{2.0cm}}
\hline
\textbf{Benchmark} & \textbf{Metric} &  \textbf{Threshold} \\
\hline
Reference Geometry Stability & RMSD (\AA) & 0.075 \cite{head-gordon2016}\\
\hline
Non-covalent Interactions & Absolute deviation from reference interaction energy (kcal/mol) & 1.0 \cite{head-gordon2016} \\
\hline
Dihedral Scan & Mean barrier error (kcal/mol) & 1.0 \cite{Boothroyd2023Sage}\\
\hline
Conformer Selection & MAE (kcal/mol) & 0.5 \\
 & RMSE (kcal/mol) & 1.5 \cite{roitberg2017}\\
\hline
Tautomers & Absolute deviation ($\Delta G$) & 0.05\\
\hline
Ring Planarity & Deviation from plane (\AA) & 0.05 \cite{Evans2007StereochemicalRestraints}\\
\hline
Bond Length Distribution & Avg. fluctuation (\AA) & 0.05 \cite{head-gordon2016} \\
\hline
Reactivity-TST & Activation Energy (kcal/mol) & 3.0 \cite{bursch2022}\\
& Enthalpy (kcal/mol) & 2.0 \cite{bursch2022}\\
\hline
Reactivity-NEB & Final force convergence (eV/\AA) & 0.05 \cite{ORCA_manual_NEB}\\
\hline
Radial Distribution Function & RMSE (\AA) & 0.1 \cite{dellago2016}\\
\hline
Protein Sampling Outliers & Ramachandran ratio & 0.1 \\
& Rotamers Ratio & 0.03 \\
\hline
Protein Folding Stability & min(RMSD) (\AA) & 2.0 \\
& max(TM-Score) & 0.5 \\
\hline
\end{tabular}
\label{tab:benchmark-acceptable-ranges}
\end{table}
\section{MLIPAudit Leaderboard Explained}
To showcase the comparative insights enabled by MLIPAudit, we have evaluated the performance of the three graph-based MLIPs provided in the open-source mlip library \cite{mlip-library2025}: MACE \cite{batatia2022mace}, NequIP \cite{batzner2022e3}, and ViSNet \cite{wang2024enhancing}. All three models were trained on a subset of the SPICE2 dataset \cite{spice-original}, which includes 1,737,896 molecular structures across 15 elements (B, Br, C, Cl, F, H, I, K, Li, N, Na, O, P, S, Si). From now on, MACE-SPICE2, NequIP-SPICE2 and ViSNet-SPICE2. Training protocols and dataset curation details are available in \cite{mlip-library2025}. We trained versions of each of these models (MACE-t1x, NequIP-t1x, ViSNet-t1x) using $10\%$ (randomly sampled) of the original t1x dataset \cite{t1x2022}, containing a total of one million structures and four elements (H, C, N, O). Additionally, we have trained two versions of ViSNet using different subsets of SPICE2 and 1 million datapoints from t1x (both taken from the OpenMolecules dataset - OMOL \cite{levine_open_2025}) , respectively, ViSNet-SPICE2(charged)-t1x, ViSNet-SPICE2(neutral)-t1x (When not specified, the neutral version is used). The mlipaudit library also supports Torch-based models as long as they have have been wrapper in an ASE Calculator class \cite{ase-paper, ISI:000175131400009}. For completeness, we have evaluated a non-exhaustive subset of Torch-based models using their original implementation, namely: MACE-OFF \cite{kovacs2025maceoff}, MACE-MP \cite{batatia2022mace}, and UMA-Small \cite{brandon-uma2025}. Two comments on these are worth raising: (1) runtime are not optimal for these models as they rely on ASE instead of JAXMD for simulations, (2) MACE-MP is trained for materials and at a different level of DFT theory. It is therefore not well suited for the benchmarks presented in MLIPAudit. We nonetheless added it as it is largely considered a reference model in the community and as results provide some interesting insights.

In Appendix \ref{appendix:sup-fig-tab}-Table \ref{tab:test-train-overlap}, we disclose the overlap between the MLIPAudit test cases per benchmark and the training set for the presented internal models. In most cases, the overlap is either zero or under 10 $\%$. But, for the conformer selection benchmark, for which two molecules (adenosine and efivarez) from the Wiggle150 \cite{wiggle150} dataset were present in the model's training set. We do not provide this information for external open source models. In the following, we will discuss the different scores and how the overlap may impact ranking.

\subsection{Overall ranking}
Table \ref{tab:overall-ranking} highlights the generalisation capabilities of the top-performing models. In the following, we will analyse separately external open-source models run using the original implementation from our internal models. Some models did not complete all benchmarks; we refer you to Appendix \ref{appendix:sup-fig-tab}, Table \ref{tab:single-rankings} for more information. Missing benchmarks can be due to the availability of elements in the training set (essentially the models trained on t1x only) or runtime issues due to the reliance of external models on ASE \cite{ase-paper, ISI:000175131400009}.

\begin{table}[h!]
\centering
\caption{Overall MLIPAudit scores}
\begin{tabular}{c|c|c|c|c|c}
\hline
\textbf{Source} & \textbf{Rank} & \textbf{Model Name} & \textbf{Average Score} & \textbf{Benchmarks} \\
\hline
External & 1 & UMA-Small & 0.70 & 12/14 \\
External & 2 & MACE-OFF & 0.63 & 11/14 \\
External & 3 & MACE-MP & 0.41 & 9/14 \\
\hline
Internal & 1 & ViSNet-SPICE2 & 0.70 & 14/14 \\
Internal & 2 & NequIP-SPICE2 & 0.70 & 14/14 \\
Internal & 3 & ViSNet-SPICE2-t1x & 0.70 & 14/14 \\
Internal & 4 & MACE-SPICE2 & 0.63 & 14/14 \\
Internal & 4 & NequIP-t1x & 0.10 & 4/14 \\
Internal & 5 & MACE-t1x & 0.10 & 4/14 \\
Internal & 6 & ViSNet-t1x & 0.10 & 4/14 \\
\hline
\end{tabular}
\label{tab:overall-ranking}
\end{table}

For the external models, UMA-Small achieves the highest average score (0.70), completing 12/14 benchmarks, followed by MACE-OFF (0.63), completing 11/14 benchmarks. MACE-MP completes 9/14 and scores 0.41; we include this model on purpose as a test for the Physics the benchmarks, as MACE-MP is trained the MPtrj dataset \cite{Deng2023} and therefore specialised on crystalline matter and not condensed matter. All internal models completed the 14 benchmarks. ViSNet-SPICE2-t1x and ViSNet-SPICE2 attain the strongest performance (0.70), closely followed by NequIP-SPICE2 (0.68) and MACE-SPICE2 (0.63). The models specifically trained on the t1x dataset \cite{t1x2022} score lower (0.1) and cover only a subset of benchmarks (4/14), reflecting the impact of training data breadth and domain coverage. Models consistently performing well across domains underscore the benefits of comprehensive training and robust architectures. However, it is worth noting that model performance is reflective of training strategy, not solely the model architecture, and it should not be considered an assessment of the model architecture. It is also important to note that UMA-Small, MACE-OFF, and MACE-MP may include train–test overlaps, and therefore their scores could be artificially overstated.

\subsection{Categorical ranking}
In Appendix \ref{appendix:sup-fig-tab}-Table \ref{tab:category-rankings}, we summarise the category-based ranking analysis, which further reveals the specialisation and limitations of each MLIP model across different chemical domains. In the General category, which tests for molecular dynamics stability, most models (internal and external) achieve perfect scores, indicating strong stability for different chemical entities in vacuum and in solution \ref{tab:test-cases}. 
The picture becomes more differentiated in the Small-molecule benchmarks. For the external models,  UMA-Small leads with a score of 0.56, followed by MACE-OFF (0.50) and MACE-MP (0.36). The ViSNet-SPICE2-t1x variant is the best internal model in this category (0.65). Among models trained purely on SPICE2 \cite{spice-original}, ViSNet-SPICE2, NequIP-SPICE2, and MACE-SPICE2 cluster closely together (0.52-0.51), demonstrating consistent performance across gas-phase and conformational tasks. In contrast, models trained primarily on the t1x dataset \cite{t1x2022} exhibit lower performance (0.11-0.16), consistent with the dataset’s focus on reactive gas-phase chemistry rather than diverse molecular energetics or equilibrium conformational distributions.
The Molecular-liquids category shows the strongest overall spread. Within the external models, UMA-Small achieves the highest score (0.98), followed by MACE-OFF (0.73). MACE-MP, trained on inorganic crystal trajectories, underperforms here (0.45), reflecting the domain shift between crystalline materials and molecular liquids. The internal models trained on SPICE2 perform similarly with scores around 0.95-0.97. These results highlight that SPICE2-trained models, despite being built from largely gas-phase and small-molecule electronic-structure data, still transfer effectively to condensed-phase structure and energetics. 
Performance diverges further in the Biomolecule category, which probes larger solvated, flexible, and chemically complex systems. External and Internal models (except for models trained exclusively on t1x) score very high in this category, around 0.8-1.0. However, MACE-MP also scores high (0.79), which highlights that the length of the simulation is not enough to assess the dynamical behaviour of the systems. Simulation length is constrained by computational resources, as this is the most expensive benchmark to run (more details will follow). t1x-trained models again unsurprisingly trail behind, consistent with their lack of exposure to biomolecular chemistry.
Overall, these results emphasise the importance of both training data diversity and domain alignment for robust generalisation across molecular and biomolecular environments, while also pointing to meaningful architectural and training-strategy differences even within closely related model families.

\subsection{Single benchmark highlighted results}
\subsubsection{Reactivity benchmarks}
Internal models trained exclusively on SPICE2 (ViSNet-SPICE2, NequIP-SPICE2, MACE-SPICE2) perform notably badly in the reactivity task with scores below 0.1 (Table \ref{tab:reactivity-benchmarks}). It is worth noting that all internal models completed all test cases (100/100 for the nudge elastic band (NEB) benchmark, $\sim$12000/12000 for the transition-state-theory (TST) benchmark), indicating that performance differences stem from modelling accuracy rather than lack of elements in the training set. These results suggest that, in the context of reactivity benchmarks, domain-specific training still offers a measurable edge, especially when accurate prediction of reaction energies or barriers is the primary objective. t1x trained models perform better in this category with scores ranging from 0.4-0.8 in the TST benchmark and 0.38-0.58 in the nudge-elastic-band (NEB) convergence benchmark, with most notably the ViSNet-SPICE2-t1x (charged and neutral) lead this category with 0.8 and 0.58, respectively. 

\begin{table}[!h]
\centering
\small
\caption{Reactivity Benchmarks Ranking}
\label{tab:reactivity-benchmarks}
\begin{tabular}{c|c|c|c|c|c}
\hline
\textbf{Source} & \textbf{Rank} & \textbf{Benchmark} & \textbf{Model Name} & \textbf{Score} & \textbf{Test Cases} \\
\hline
External & 1& Small Molecule Reactivity TST & UMA-Small & 0.86 & 11961/11961 \\
External & 2 & Small Molecule Reactivity TST & MACE-OFF & 0.12 & 11961/11961 \\
External & 3 & Small Molecule Reactivity TST & MACE-MP & 0.05 & 11961/11961 \\
\hline
Internal & 1 & Small Molecule Reactivity TST & ViSNET-SPICE2-t1x & 0.77 & 11961/11961 \\
Internal & 2 & Small Molecule Reactivity TST & NequIP-t1x & 0.41 & 11961/11961 \\
Internal & 3 & Small Molecule Reactivity TST & MACE-t1x & 0.39 & 11961/11961 \\
Internal & 3 & Small Molecule Reactivity TST & ViSNET-t1x & 0.39 & 11961/11961 \\
Internal & 4 & Small Molecule Reactivity TST & MACE-SPICE2 & 0.1 & 11961/11961 \\
Internal & 5 & Small Molecule Reactivity TST & ViSNET-SPICE2 & 0.05 & 11961/11961 \\
Internal & 5 & Small Molecule Reactivity TST & NequIP-SPICE2 & 0.05 & 11961/11961 \\
\hline
Internal & 1 & Small Molecule Reactivity NEB & ViSNET-SPICE2-t1x & 0.58 & 100/100 \\
Internal & 2 & Small Molecule Reactivity NEB & NequIP-t1x & 0.58 & 100/100 \\
Internal & 3 & Small Molecule Reactivity NEB & MACE-t1x & 0.44 & 100/100 \\
Internal & 3 & Small Molecule Reactivity NEB & ViSNET-t1x & 0.38 & 100/100 \\
Internal & 4 & Small Molecule Reactivity NEB & MACE-SPICE2 & 0.1 & 100/100 \\
Internal & 4 & Small Molecule Reactivity NEB & ViSNET-SPICE2 & 0.1 & 100/100 \\
Internal & 4 & Small Molecule Reactivity NEB & NequIP-SPICE2 & 0.1 & 100/100 \\
\hline
\end{tabular}
\end{table}
\begin{figure}[!h]
    \centering
    \includegraphics[width=0.95\linewidth]{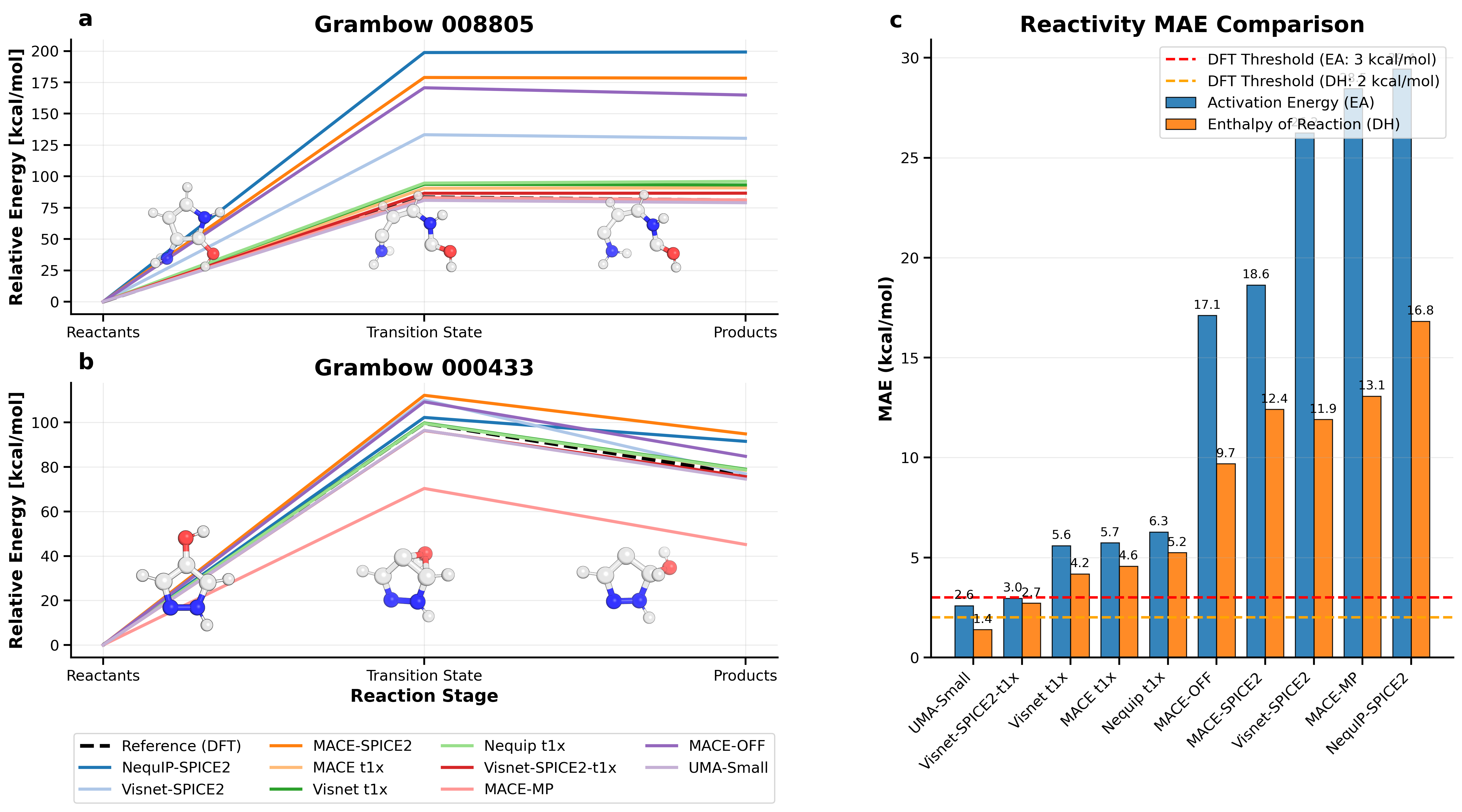}
    \caption{Reactivity benchmark performance.
(a–b) Reaction energy profiles for two Grambow reactions (IDs 008805 and 000433) \cite{grambow} MLIP predictions to DFT references. (c) MAEs for activation energies (EA) and reaction enthalpies across the benchmark.}
    \label{fig:reactivity-example}
\end{figure}

As shown in Figure \ref{fig:reactivity-example}, all t1x-trained models outperform SPICE2 trained MLIPs (and SPICE1 in the case of MACE-OFF), which show much larger errors, especially for activation energies.

From the external models, UMA-Small excels in the reactivity benchmark with a score of 0.86, with MACE-OFF following behind with a score of 0.12. While remarkable, all our test-cases come from the Grambow dataset \cite{grambow}, which is included in the t1x dataset \cite{t1x2022}, which is included in full in the UMA-Small training data.

\subsubsection{Molecular liquids benchmark: water radial distribution function}
\begin{figure}[!h]
    \centering
    \includegraphics[width=0.9\linewidth]{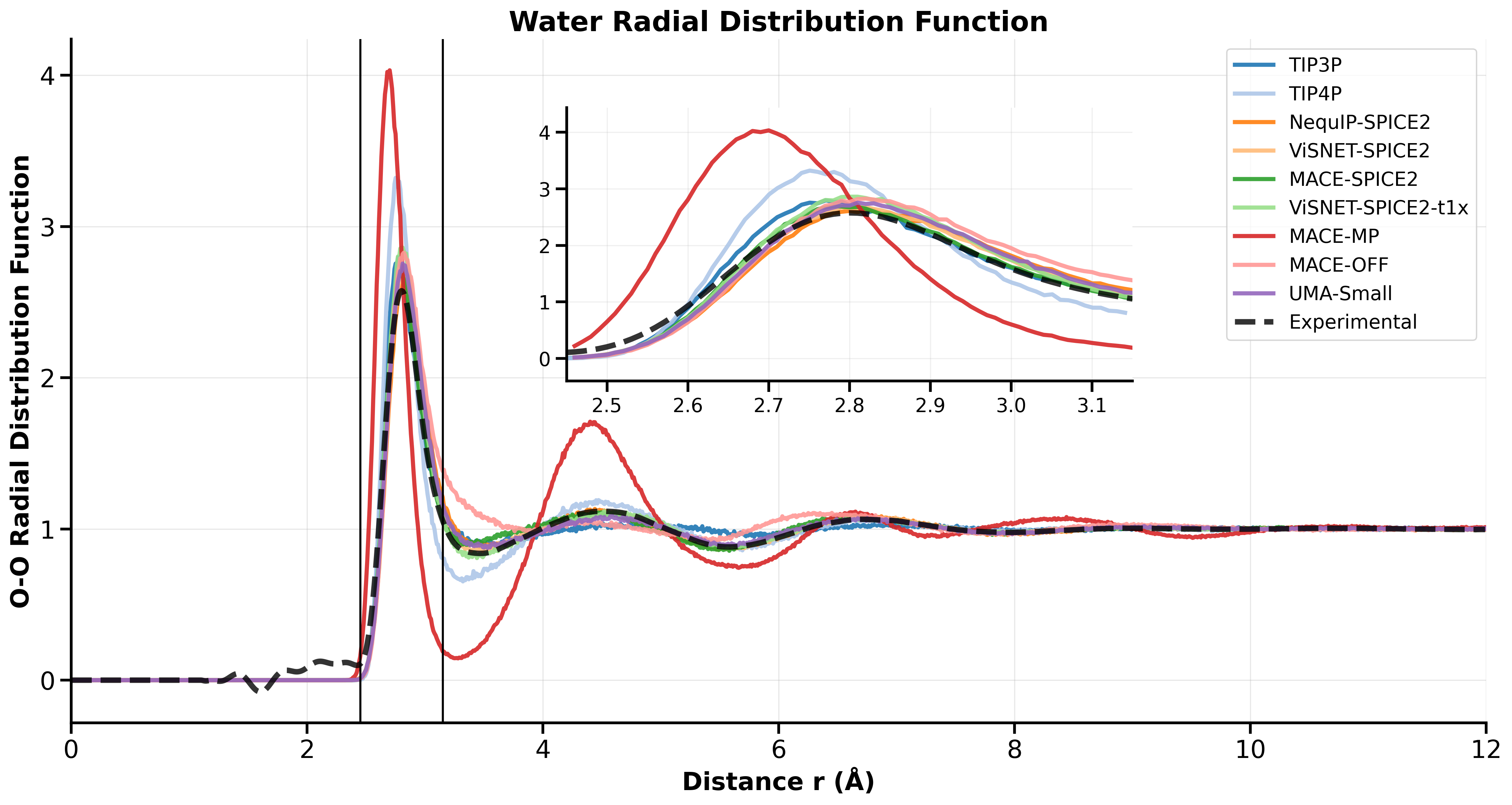}
    \caption{Water radial distribution function for the example models, compared with the experimental observable and two water classical forcefields TIP3P and TIP4P \cite{tip3p}}
    \label{fig:water-rdf}
\end{figure}

Having a closer look at the single benchmarks, the water radial distribution function (RDF) benchmark provides a compelling illustration of the strengths of MLIPs over traditional force fields. As shown in Figure \ref{fig:water-rdf}, all five internal MLIP models, MACE-SPICE2, ViSNet-SPICE2, ViSNet-SPICE2(neutral)-t1x, ViSNet-SPICE2(charged)-t1x, ViSNet-SPICE2 and NequIP-SPICE2, reproduce the experimental RDF profile with high fidelity across the full radial range, accurately reproducing both the first solvation shell peak and subsequent oscillations. And this is also true for the original implementations of UMA-Small and MACE-OFF. In contrast, TIP3P and TIP4P \cite{tip3p}, two of the most widely used classical water models, show notable deviations, particularly in the overstructured and exaggerated height of the first peak, a known artefact in rigid water models \cite{Camisasca2019}. 
Notably, MACE-MP produces crystalline water even when simulated at 300 K, indicating that the model remains strongly biased toward its crystal-structure training data despite liquid-phase simulation conditions. This behaviour is evident in the radial distribution function (RDF): whereas liquid water shows a broadened first O–O peak near ~2.8 Å and damped oscillations characteristic of short-range order, crystalline (ice-like) water exhibits sharp, well-defined peaks extending to long range, reflecting persistent translational order. These qualitative differences are well-established in the literature \cite{Skyner2017WaterHydrates}.

This alignment between MLIP predictions and experimental data strongly supports the notion that learned potentials, trained on accurate quantum data, can capture the subtle balance of hydrogen bonding and thermal fluctuations that define liquid water structure, without the need for hand-tuned parametrisation. This not only reflects the higher representational capacity of MLIPs but also demonstrates their ability to generalise to bulk-phase properties, a capability that classical force fields struggle to match without introducing complex polarisable terms or many-body corrections.

\subsubsection{Small molecules benchmarks: dihedral scans}
\begin{figure}[!h]
    \centering
    \includegraphics[width=0.95\linewidth]{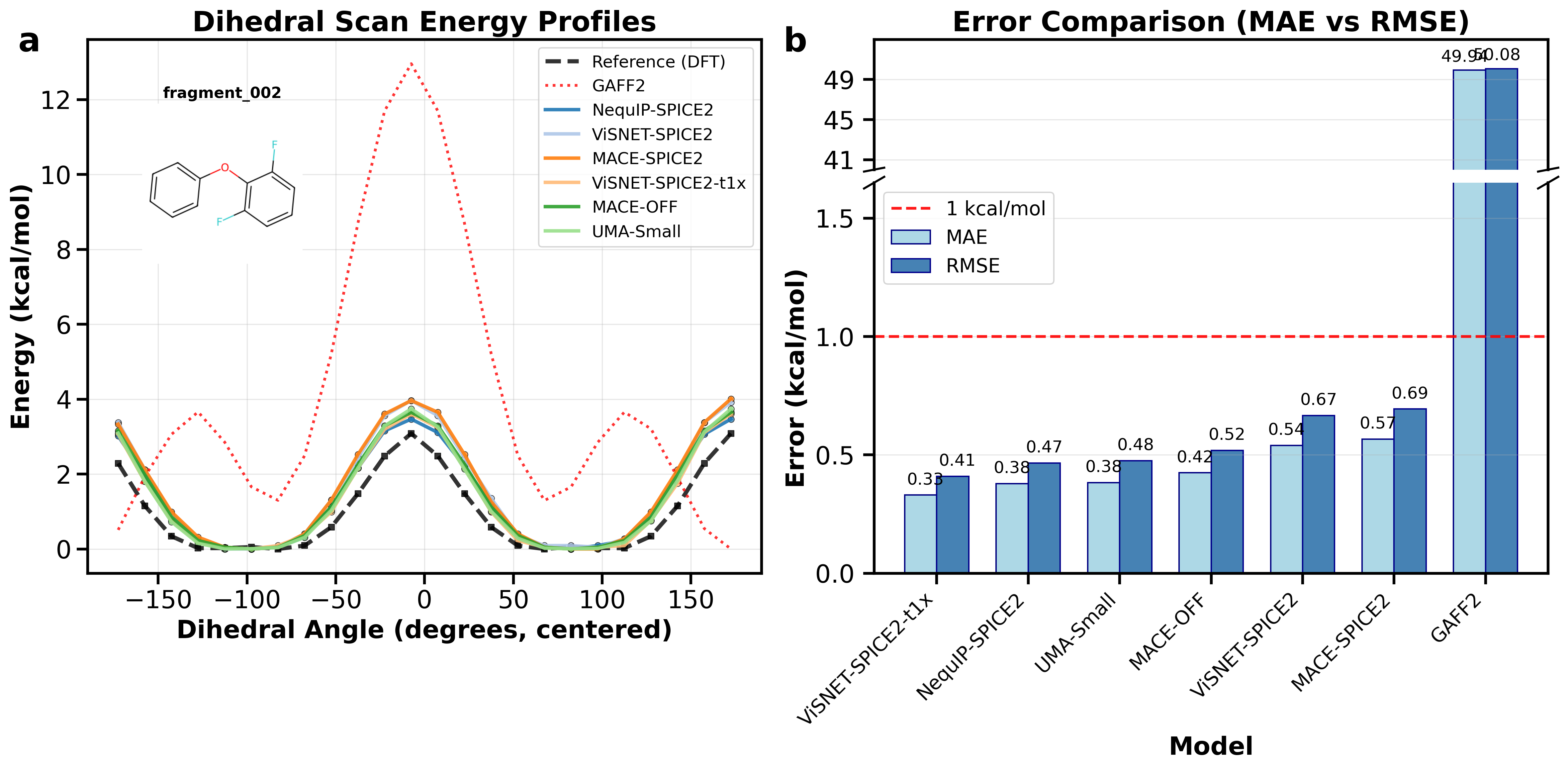}
    \caption{Dihedral scan benchmark.
(a) Dihedral energy profiles for fragment 015 compared to DFT reference values. (b) MAE and RMSE for each model. DFT-level error threshold (red dashed line).}
    \label{fig:dihedral-scan}
\end{figure}

The dihedral scan benchmark highlights another area where MLIP models show outstanding agreement with quantum reference data. As shown in Figure \ref{fig:dihedral-scan}, the energy profiles predicted by all MLIP models align nearly perfectly with DFT-calculated torsional energy curves across a representative scan. This agreement is not only qualitative—preserving the positions and heights of barriers, but also quantitatively precise, with RMSE values all well below the 1.0 kcal/mol DFT-level convergence threshold.
This strong performance is further reflected in the ranking table (Appendix \ref{appendix:sup-fig-tab}, Table  \ref{tab:category-rankings}), where ViSNet-SPICE2 and ViSNet-SPICE2-t1x lead the benchmark scoring $\sim$1.0, followed closely by NequIP-SPICE2 and MACE-SPICE2, MACE-SPICE2-t1x. Notably, all models completed the full set of 500 fragments, demonstrating not only accuracy but robustness and generalisability across a diverse chemical space.

The error bars shown on the right panel of Figure \ref{fig:dihedral-scan} underscore how consistent the models are, with MAE values under 0.12 kcal/mol for all methods—well within chemical accuracy. MLIPs outperform classical parameters like GAFF2 \cite{gaff2}. These results validate the capability of current MLIPs to accurately model intramolecular potential energy surfaces, a critical requirement for reliable conformational sampling, molecular docking, or pharmacophore prediction.

Taken together, this benchmark provides a clear example of how MLIPs can match DFT accuracy at a fraction of the computational cost, making them practical for high-throughput screening or molecular simulations involving flexible, drug-like molecules.

\subsubsection{Small molecules benchmarks: conformer ranking }
\begin{figure}[!h]
    \centering
    \includegraphics[width=0.95\linewidth]{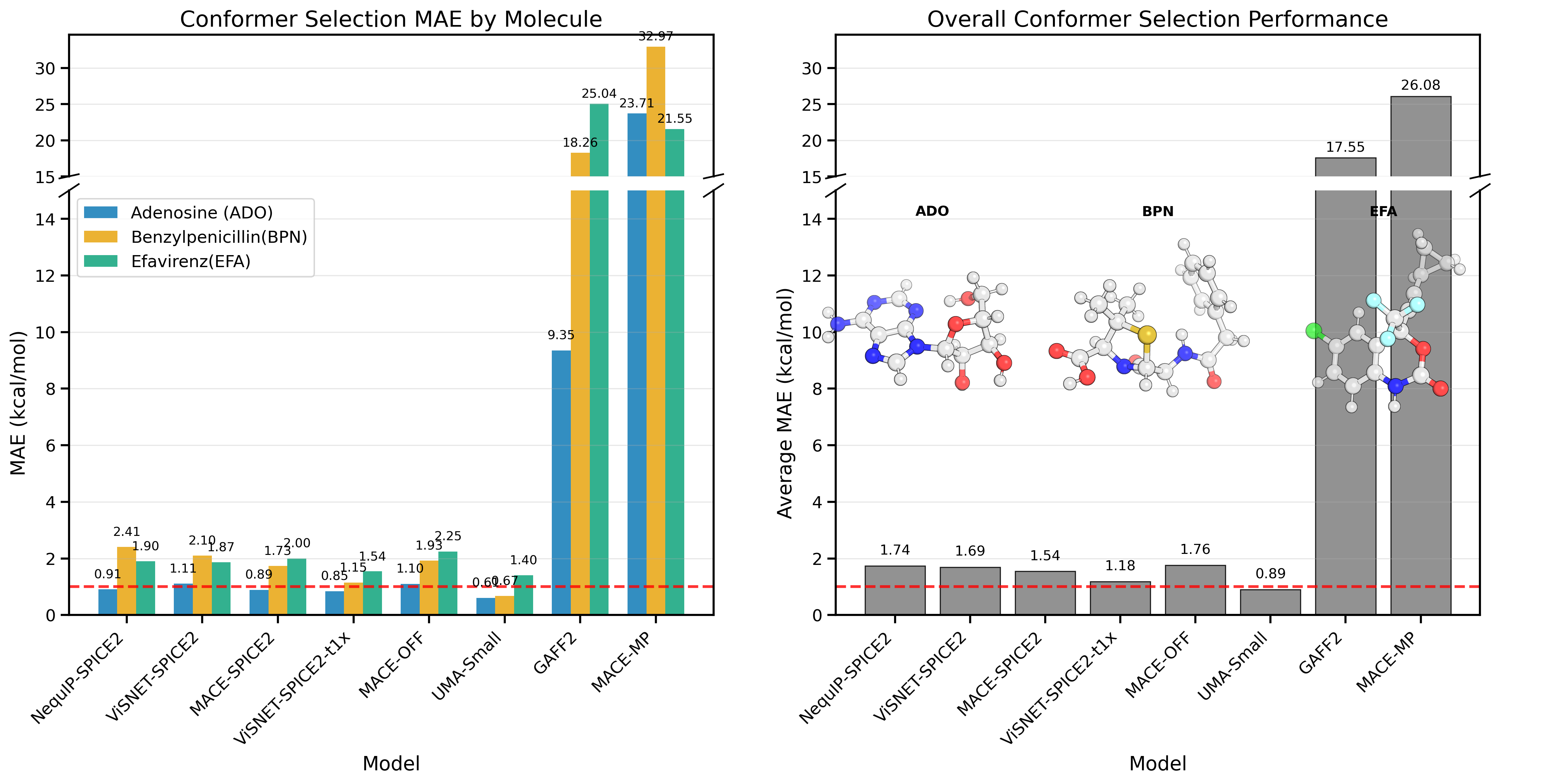}
    \caption{Conformer selection benchmark across three pharmaceutically relevant molecules: adenosine (ADO), benzylpenicillin (BPN), and efavirenz (EFA). MAE is computed with respect to DFT reference conformer energies. DFT threshold (red dashed line at 0.5 kcal/mol). Insets depict representative 3D conformers for each molecule.}
    \label{fig: conformer}
\end{figure}

Figure \ref{fig: conformer} presents model performance on the conformer benchmark, showing MAE values by molecule for three general-purpose MLIPs: NequIP-SPICE2, ViSNet-SPICE2, and MACE-SPICE2. All models were trained on datasets that included adenosine (ADO) and efavirenz (EFA), while benzylpenicillin (BPN) was excluded from training and thus acts as a stronger generalisation test.
Despite having seen ADO and EFA during training, none of the models reach the DFT-level MAE threshold of 0.5 kcal/mol, pointing to persistent difficulty in accurately ranking conformers. ADO is best predicted, while EFA shows higher errors due to its flexibility. BPN, which was unseen during training, is the most challenging, though MACE-SPICE2 shows slightly better generalisation. All models outperform GAFF2 \cite{gaff2}, especially on EFA. Still, as seen in Figure \ref{fig: conformer}, predicted vs. DFT energy plots show strong agreement and near-perfect Spearman correlations across all molecules.

This consistency suggests that while the models may struggle to reproduce exact conformer energy magnitudes (as seen in the MAE analysis), they are highly effective at preserving the correct energetic ordering. In practical applications like conformer selection or ranking, such ordinal accuracy can be more important than precise energetic reproduction, particularly when used in combination with scoring functions or downstream screening.

Interestingly, the performance gap between in-training-set molecules (ADO, EFA) and the out-of-distribution case (BPN) is far less pronounced here than in absolute MAE terms—highlighting that model generalisation, at least in terms of correlation, is relatively robust. These findings reinforce the importance of using multiple complementary metrics (e.g., MAE and rank correlation) when evaluating MLIP performance for conformational energetics.

\subsubsection{Biomolecules benchmarks}
The biomolecules benchmark (Appendix \ref{appendix:sup-fig-tab}, Table \ref{tab:category-rankings}) provides a fitting conclusion to our comprehensive assessment, highlighting the potential for MLIP models to operate effectively in complex, biologically relevant regimes. The biomolecules benchmark is the most computationally intensive one, as it involves solvated systems with 1000 to 4000 atoms in total (Appendix \ref{appendix:sup-fig-tab}, Table \ref{tab:test-cases}). 

All top models successfully completed the protein folding stability benchmark (3/3 test cases, see Appendix \ref{appendix:sup-fig-tab}), all models achieve similar scores $\sim$0.525, but there is room for improvement. This level of agreement underscores the growing maturity of MLIPs for macromolecular tasks. The Protein Sampling benchmark across different MLIP models shows that models trained on the SPICE2 dataset (e.g., ViSNet-SPICE2, NequIP-SPICE2, MACE-SPICE2) significantly outperform their t1x-trained counterparts, with ViSNet-SPICE2 achieving the highest score (0.928) and full coverage (12/12 systems). Taken together, the results from this and all previous benchmarks reinforce a central conclusion: while task-specific training offers advantages in specialised domains, the leading models demonstrate strong, transferable performance across molecular scales and properties, setting the stage for robust deployment in real-world chemistry and biology applications.
\section{Conclusions and future outlook}
The MLIPAudit suite provides a comprehensive and diverse evaluation framework for MLIPs, spanning small-molecule geometrical and conformational energetics, reactivity, molecular liquids, and biomolecular stability and sampling. Results show that while specialised models trained on the t1x dataset excel in targeted tasks such as reaction barrier prediction, general-purpose architectures like ViSNet-SPICE2, NequIP-SPICE2, and MACE-SPICE2 exhibit strong and transferable accuracy across a wide range of benchmarks, often surpassing classical force fields and closely matching DFT reference data in others. Notably, the ViSNet model trained on SPICE2 and t1x from the OMOL dataset leads the small-molecule benchmarks, highlighting the promise of hybrid training strategies and possibly reflecting the importance of the underlying level of theory used in data generation.

Despite this progress, performance gaps persist, especially in condensed-phase systems and energetically subtle regimes, indicating that further improvements are needed. While MLIPAudit establishes a unified and reproducible evaluation suite, it also has limitations. The current set of models is biased toward graph neural network architectures, and the benchmarks rely primarily on DFT data of varying origin, which may introduce systematic bias. Efficiency and robustness-oriented metrics (e.g., uncertainty calibration and scalability) are not yet fully assessed, and several critical chemical regimes, such as transition-metal systems, enzyme catalysis, and extreme thermodynamic conditions, remain under-represented due to limited reference data.

A further challenge lies in maintaining truly blind test sets. As the community continually expands training datasets, ensuring that future benchmark systems remain unseen becomes increasingly difficult. In future iterations, we will explore generating dedicated blind datasets and curated QM reference sets, though this task will remain increasingly complex.

Future releases will introduce more demanding simulation tasks, such as free-energy estimation, reactive condensed-phase processes, and protein–ligand systems. By evolving alongside the MLIP community and enabling continuous contribution, MLIPAudit aims not only to benchmark progress but to support rigorous, open, and scalable development of next-generation ML interatomic potentials. By continually broadening the scope and complexity of MLIPAudit, we hope to accelerate the development of MLIPs that are not only accurate but also general, scalable, and ready for real-world deployment across the chemical sciences.
\section{Known scope gap of the benchmarks}
Although MLIPAudit provides a diverse collection of benchmarks spanning general MD stability, small-molecule energetics, molecular liquids, and biomolecular systems, several scope limitations remain in this initial release. These gaps reflect the practical constraints of reference data availability, the computational cost of current MLIPs, and the ambition to balance breadth with tractability.

First, the benchmark suite does not yet capture the full complexity of chemical and biophysical environments. The benchmarks currently include:

General: basic simulation stability;

Small molecules: reference-geometry stability, conformer ranking, dihedral scans, non-covalent interactions, transition-state and NEB reaction profiles, tautomers, and geometric planarity;

Liquids: water radial and angular distribution functions and solvent radial distributions;

Biomolecules: folding stability and local protein sampling.

While this coverage provides a representative cross-section of molecular and condensed-phase challenges, many essential regimes remain outside the present scope. Notably, heterogeneous interfaces, extended membrane–protein systems, reactive events, electronically excited states, and large-scale conformational processes (e.g., folding pathways, allosteric transitions) are not assessed. These phenomena either require reference datasets that are not yet widely available or involve quantum-mechanical costs prohibitive for large-scale benchmarking.

Second, although biomolecular benchmarks are performed in explicit water, their temporal extent is necessarily limited. Current MLIP implementations remain too computationally expensive to routinely support long-timescale simulations for large solvated proteins. As a result, biomolecular MD benchmarks in MLIPAudit are restricted to short (250 ps) trajectories. These simulations are long enough to detect major instabilities, force noise, or poor sampling behaviour, but do not capture slower biophysical processes such as large-amplitude rearrangements, side-chain relaxation, or folding transitions. Consequently, the biomolecular category should be viewed as an early stability and local-sampling diagnostic rather than a full assessment of MLIP performance on biologically relevant timescales.

Finally, important physical regimes — including high-pressure or extreme-temperature conditions, far-from-equilibrium dynamics, chemical reactivity, and solid-state properties — are intentionally not covered in this release. Users interested in inorganic crystalline behaviour (e.g., phonons, defects, phase transitions) may need to complement MLIPAudit with existing materials-focused benchmarks to obtain a complete evaluation.

These scope gaps reflect the intentional design of MLIPAudit as a modular and extensible benchmark suite. Future community contributions and the continued evolution of MLIP methods will enable richer datasets, broader chemical coverage, and more demanding dynamical evaluations in subsequent releases.
\section{Acknowledgements}
This work was supported by Cloud TPUs from Google’s TPU Research Cloud (TRC).

\section{Appendix}
\subsection{Model training details}

Three of the models presented in this paper were released as part of the mlip library \cite{mlip-library2025}: ViSNet-SPICE2, MACE-SPICE2, and NequIP-SPICE2. Details on how these models were trained, alongside training data details and hyperparameters can be found in the original reference. 

We present in Table \ref{tab:model-training-details} below details about the other models presented as examples in the paper. Note that training details for UMA-Small, MACE-OFF and MACE-MP
can be found in Ref. \cite{brandon-uma2025}, \cite{kovacs2025maceoff}, and \cite{batatia2025foundationmodelatomisticmaterials} respectively. 

\begin{table}[!h]
\centering
\small
\caption{Example models training details}
\begin{tabular}{l|p{5.5cm}|p{4cm}}
\hline
\textbf{Model} & \textbf{Dataset} &  \textbf{Hyperparameters} \\
\hline
ViSNet-SPICE2 & Original version of SPICE2 \cite{spice-original}, as curated in \cite{mlip-library2025} - includes only neutral systems & As described in \cite{mlip-library2025} \\
\hline
MACE-SPICE2 & Original version of SPICE2 \cite{spice-original}, as curated in \cite{mlip-library2025} - includes only neutral systems & As described in \cite{mlip-library2025} \\
\hline
NequIP-SPICE2 & Original version of SPICE2 \cite{spice-original}, as curated in \cite{mlip-library2025} - includes only neutral systems & As described in \cite{mlip-library2025} \\
\hline
ViSNet-t1x & Original version of Transition-1X \cite{t1x2022}, trained on 1M samples, randomly sampled with 95/5 train/val split. & Same as ViSNet-SPICE2 \\
\hline
MACE-t1x & Original version of Transition-1X \cite{t1x2022}, trained on 1M samples, randomly sampled with 95/5 train/val split. & Same as MACE-SPICE2\\
\hline
NequIP-t1x & Original version of Transition-1X \cite{t1x2022}, trained on 1M samples, randomly sampled with 95/5 train/val split. & Same as NequIP-SPICE2\\
\hline
ViSNet-SPICE2(charged)-t1x  & SPICE2 and Transition-1X as recomputed in the OMOL dataset \cite{levine_open_2025}. SPICE2 is curated as is described in \cite{mlip-library2025}. T1X includes 50k samples, selected among transition states, reactants and products. & Same as ViSNet-SPICE2, except for the number of channels with is increased to 256. \\
\hline
ViSNet-SPICE2(neutral)-t1x  & SPICE2 and Transition-1X as recomputed in the OMOL dataset \cite{levine_open_2025}. SPICE2 is curated as is described in \cite{mlip-library2025}. T1X includes 1M samples, selected among transition states, reactants and products. & Same as ViSNet-SPICE2, except for the number of channels with is increased to 256. \\
\hline
\end{tabular}
\label{tab:model-training-details}
\end{table}

\subsection{Supporting Figures and Tables}
\label{appendix:sup-fig-tab}
\begin{table}[!h]
\centering
\small
\caption{Datasets used for the different benchmarks in MLIPAudit.}
\label{tab:test-cases}
\begin{tabular}{l|p{3cm}|p{3.2cm}|p{4.5cm}}
\hline
\textbf{Benchmark} & \textbf{Dataset Name} & \textbf{Link/Citation} & \textbf{Content Description} \\
\hline
General Stability & In-house dataset & Released with MLIPAudit & 3 small molecules in vacuum (1 HCNO-only, 1 with halogens, 1 with sulfur). 2 peptides in vacuum (Neurotensin PDBid 2LNF and Oxytocin PDBid 7OFG). 1 protein in vacuum (PDBid 1A7M). 1 peptide in pure water (Oxytocin). 1 peptide in water with Cl- counterions (Neurotensin).\\
\hline
Inference Scaling & In-house dataset & Released with MLIPAudit & Proteins in vacuum. PDBids: 1AY3, 1UAO, 1AB7, 1P79, 1BIP, 1A5E, 1A7M, 2BQV, 1J7H, 5KGZ, 1VSQ, 1JRS. \\
\hline
Reference Geometry Stability & OpenFF & \cite{openFF-dataset} & 200 molecules for the neutral dataset and 20 for the charged dataset. The subsets are constructed so that the chemical diversity, as represented by Morgan fingerprints, is maximised. \\
\hline
Non-covalent Interactions & NCI-ATLAS subsets: \newline
D442x10, HB375x10, \newline
HB300SPXx10, \newline
IHB100x10, R739x5, \newline
SH250x10 & http://www.nciatlas.org/ & QM optimised geometries of distance scans of bi-molecular complexes, where the two molecules interact via non-covalent interactions with associated energies. \\
\hline
Dihedral Scan & In-house recomputed TorsionNet 500 dataset at $\omega$B97M-D3(BJ) DFT-level. & \cite{torsionnet} & 500 structures of drug-like molecules and their energy profiles around selected rotatable bonds at wB97M-D3(BJ) DFT-level. \\
\hline
Conformer Selection & Wiggle 150 & \cite{wiggle150} & 50 conformers each of three molecules: Adenosine, Benzylpenicillin, and Efavirenz. \\
\hline
Tautomers & In-house recomputed Tautobase dataset at $\omega$B97M-D3(BJ) DFT-level. & \cite{tautobase} & 2,792 tautomer pairs sourced from the Tautobase dataset. After generation of the structures and minimisation at xtb level, the QM energies were computed in-house using $\omega$B97M-D3(BJ)/def2-TZVPPD level of theory. \\
\hline
Ring Planarity & QM9 subset & \cite{qm9-dataset} & One representative molecule each, containing substructures for benzene, furan, imidazole, purine, pyridine and pyrrole. \\
\hline
Bond Length & QM9 subset & \cite{qm9-dataset} & One representative molecule each, containing the bond types C-C, C=C, C$\#$C, C-N, C-O, C=O and C-F. \\
\hline
Reactivity & Grambow dataset & \cite{grambow} & Reactants, products and transition states of 11960 reactions. \\
\hline
Radial Distribution Function & In-house solvent boxes & Released with MLIPAudit. Reference data: \cite{water-rdf, ccl4, methanol,acetonitrile} & Water, CCl4, Acetonitrile, Methanol.\\
\hline
Protein Folding Stability & In-house dataset & Released with MLIPAudit & 3 solvated proteins: Chignolin, Orexin and Trp Cage. PDBids: 1UAO, 2JOF, 1CQ0.\\
\hline
\end{tabular}
\end{table}
\begin{table}[h]
\centering
\caption{MLIPAudit test-cases overlap with models training dataset for internal models only}
\label{tab:test-train-overlap}
\newpage
\begin{tabular}{l|l|c}
\hline
\textbf{Benchmark Category} & \textbf{Benchmark} & \textbf{Overlap [\%]} \\
\hline
Small-Molecule & Reference Geometry Stability & 0 \\
Small-Molecule & Bond Length distribution & 0 \\
Small-Molecule & Ring Planarity & 0 \\
Small-Molecule & Conformer selection & 66.7 \\
Small-Molecule & Dihedral scan & 1.4 \\
Small-Molecule & Tautomers & 8.4 \\
Small-Molecule & Non-covalent interactions & -- \\
Small-Molecule & Reactivity & -- \\
\hline
Molecular liquids & RDF & 0 \\
\hline
Biomolecules & Folding stability & 0 \\
Biomolecules & Sampling & 0 \\
\hline
\end{tabular}
\end{table}
\newpage
\begin{table}[h!]
\centering
\caption{Category-based rankings (aggregated scores by benchmark category)}
\begin{tabular}{c|c|c|c|c|c|c}
\hline
\textbf{Source} & \textbf{Rank} & \textbf{Category} & \textbf{Model Name} & \textbf{Score} & \textbf{Metrics} \\
\hline
External & 1 & General & UMA-Small & 1.00 & 1/1 \\
External & 1 & General & MACE-SPICE2 & 1.00 & 1/1 \\
External & 1 & General & MACE-MP & 1.00 & 1/1 \\
\hline
Internal &1 & General & ViSNet-SPICE2 & 1.00 & 1/1\\
Internal &1 & General & MACE-SPICE2 & 1.00 & 1/1 \\
Internal &2 & General & NequIP-SPICE2 & 0.90 & 1/1 \\
Internal & 3 & General & ViSNet-SPICE2-t1x & 0.75 & 1/1\\
Internal &3 & General & ViSNet-t1x & 0.00 & 1/1 \\
Internal &3 & General & NequIP-t1x & 0.00 & 1/1 \\
Internal &3 & General & MACE-t1x & 0.00 & 1/1 \\
\hline
\hline
External & 1 & Small-molecules & UMA-Small & 0.56 & 7/9 \\
External & 2 & Small-molecules & MACE-OFF & 0.50 & 8/9 \\
External & 3 & Small-molecules & MACE-MP & 0.36 & 7/9 \\
\hline
Internal & 1 & Small-molecules& ViSNet-SPICE2-t1x & 0.65 & 9/9 \\
Internal &2 & Small-molecules & ViSNet-SPICE2& 0.52 & 9/9 \\
Internal &2 & Small-molecules & NequIP-SPICE2& 0.52 & 9/9 \\
Internal &3 & Small-molecules & MACE-SPICE2 & 0.51 & 9/9 \\
Internal &4 & Small-molecules & NequIP-t1x & 0.16 & 6/9 \\
Internal &5 & Small-molecules & MACE-t1x & 0.14 & 6/9 \\
Internal &6 & Small-molecules & ViSNet-t1x & 0.11 & 6/9 \\
\hline
\hline
External & 1 & Molecular-liquids & UMA-Small & 0.98 & 2/2 \\
External & 2 & Molecular-liquids & MACE-OFF & 0.73 & 2/2 \\
External & - & Molecular-liquids & MACE-MP & 0.0 & 2/2 \\
\hline
Internal & 1 & Molecular-liquids & NequIP-SPICE2 & 0.97 & 2/2 \\
Internal & 1 & Molecular-liquids & MACE-SPICE2 & 0.97 & 2/2 \\
Internal & 1 & Molecular-liquids & MACE-SPICE2 & 0.97 & 2/2 \\
Internal & 2 & Molecular-liquids & ViSNet-SPICE2-t1x & 0.95 & 2/2 \\
Internal & - & Molecular-liquids & ViSNet-t1x & 0.0 & 2/2 \\
Internal & - & Molecular-liquids & NequIP-t1x & 0.0 & 2/2 \\
Internal & - & Molecular-liquids & MACE-t1x & 0.0 & 2/2 \\
\hline
\hline
External & 1 & Biomolecules & UMA-Small & 0.92 & 2/2 \\
External & 1 & Biomolecules & MACE-OFF & 0.92 & 2/2 \\
External & - & Biomolecules & MACE-MP & 0.79 & 2/2 \\
\hline
Internal & 1 & Biomolecules & ViSNet-SPICE2 & 1.00 & 2/2 \\
Internal & 1 & Biomolecules & NequIP-SPICE2 & 1.00 & 2/2 \\
Internal & 2 & Biomolecules & ViSNet-SPICE2-t1x & 0.43 & 2/2 \\
Internal & - & Biomolecules & ViSNet-t1x & 0.0 & 2/2 \\
Internal & - & Biomolecules & NequIP-t1x & 0.0 & 2/2 \\
Internal & - & Biomolecules & MACE-t1x & 0.0 & 2/2 \\
\hline
\end{tabular}
\label{tab:category-rankings}
\end{table}

\begin{table}[h!]
\centering
\caption{Single benchmarks rankings}
\label{tab:single-rankings}
\begin{tabular}{c|c|c|c|c|c|c}
\hline
\textbf{Source} & \textbf{Rank} & \textbf{Benchmark} & \textbf{Model Name} & \textbf{Score} & \textbf{Test Cases} \\
\hline
External & 1 & General Stability & UMA-Small & 1.00 & 8/8 \\
External & 1 & General Stability & MACE-OFF & 1.00 & 8/8 \\
External & 1 & General Stability & MACE-MP & 1.00 & 8/8 \\
\hline
Internal & 1 & General Stability & ViSNet-SPICE2 & 1.00 & 8/8 \\
Internal & 1 & General Stability & MACE-SPICE2 & 1.00 & 8/8 \\
Internal & 2 & General Stability & Nequip-SPICE2 & 0.90 & 8/8 \\
Internal & 3 & General Stability & ViSNet-SPICE2-t1x & 0.75 & 8/8 \\
Internal & 4 & General Stability & MACE-t1x & 0.44 & 8/8 \\
\hline
External & 1 & Solvent RDF & UMA-Small & 0.95 & 3/3 \\
External & 2 & Solvent RDF & MACE-OFF & 0.73 & 3/3 \\
External & - & Solvent RDF & MACE-MP & 0.00 & 0/3 \\
\hline
Internal & 1 & Solvent RDF & Nequip-SPICE2 & 0.97 & 3/3 \\
Internal & 2 & Solvent RDF & MACE-SPICE2 & 0.94 & 3/3 \\
Internal & 2 & Solvent RDF & ViSNet-SPICE2 & 0.94 & 3/3 \\
Internal & 3 & Solvent RDF & ViSNet-SPICE2-t1x & 0.90 & 3/3 \\
\hline
External & 1 & Water RDF & UMA-small & 1.00 & 1/1 \\
External & 2 & Water RDF & MACE-OFF & 0.56 & 1/1 \\
External & - & Water RDF & MACE-MP & 0.00 & 1/1 \\
\hline
Internal & 1 & Water RDF & ViSNet-SPICE2 & 1.00 & 1/1 \\
Internal & 1 & Water RDF & MACE-SPICE2 & 1.00 & 1/1 \\
Internal & 1 & Water RDF & Nequip-SPICE2 & 1.00 & 1/1 \\
Internal & 1 & Water RDF & ViSNet-SPICE2-t1x & 1.00 & 1/1 \\
Internal & - & Water RDF & ViSNet-t1x & 0.00 & 1/1 \\
Internal & - & Water RDF & MACE-t1x & 0.00 & 1/1 \\
Internal & - & Water RDF & Nequip-t1x & 0.00 & 1/1 \\
\hline
External & 1 & Protein Folding Stability & UMA-Small & 1.00 & 3/3 \\
External & 1 & Protein Folding Stability & MACE-OFF & 1.00 & 3/3 \\
External & 1 & Protein Folding Stability & MACE-MP & 1.00 & 3/3 \\
\hline
Internal & 1 & Protein Folding Stability & ViSNet-SPICE2 & 1.00 & 3/3 \\
Internal & 1 & Protein Folding Stability & Nequip-SPICE2 & 1.00 & 3/3 \\
Internal & 2 & Protein Folding Stability & MACE-SPICE2 & 0.33 & 3/3 \\
Internal & - & Protein Folding Stability & ViSNet-SPICE2-t1x & 0.00 & 3/3 \\
\hline
\end{tabular}
\end{table}

\begin{table}[h!]
\centering
\caption{Single benchmarks rankings (cont.)}
\label{tab:single-rankings-cont}
\begin{tabular}{c|c|c|c|c|c}
\hline
\textbf{Source} & \textbf{Rank} & \textbf{Benchmark} & \textbf{Model Name} & \textbf{Score} & \textbf{Test Cases} \\
\hline
External & 1 & Reference Geometry Stability & UMA-Small & 0.98 & 220/220 \\
External & 1 & Reference Geometry Stability & MACE-OFF & 0.93 & 220/220 \\
External & 1 & Reference Geometry Stability & MACE-MP & 0.50 & 220/220 \\
\hline
Internal & 1 & Reference Geometry Stability & ViSNet-SPICE2-t1x & 0.97 & 220/220 \\
Internal & 1 & Reference Geometry Stability & ViSNet-SPICE2 & 0.97 & 220/220 \\
Internal & 2 & Reference Geometry Stability & MACE-SPICE2 & 0.96 & 220/220 \\
Internal & 3 & Reference Geometry Stability & Nequip-SPICE2 & 0.94 & 220/220 \\
\hline
External & 1 & Conformer Selection & UMA-Small & 0.29 & 3/3 \\
External & - & Conformer Selection & MACE-OFF & 0.00 & 3/3 \\
External & - & Conformer Selection & MACE-MP & 0.00 & 3/3 \\
\hline
Internal & 1 & Conformer Selection & ViSNet-SPICE2-t1x & 0.05 & 3/3 \\
Internal & 2 & Conformer Selection & Nequip-SPICE2 & 0.03 & 3/3 \\
Internal & 2 & Conformer Selection & MACE-SPICE2 & 0.03 & 3/3 \\
Internal & - & Conformer Selection & Visnet-SPICE2 & 0.00 & 3/3 \\
\hline
External & 1 & Dihedral Scan & UMA-Small & 0.71 & 500/500 \\
External & 2 & Dihedral Scan & MACE-OFF & 0.66 & 500/500 \\
External & 2 & Dihedral Scan & MACE-MP & 0.40 & 500/500 \\
\hline
Internal & 1 & Dihedral Scan & ViSNet-SPICE2-t1x & 0.70 & 500/500 \\
Internal & 2 & Dihedral Scan & ViSNet-SPICE2 & 0.69 & 500/500 \\
Internal & 2 & Dihedral Scan & Nequip-SPICE2 & 0.66 & 500/500 \\
Internal & 3 & Dihedral Scan & MACE-SPICE2 & 0.65 & 500/500 \\
\hline
External & 1 & Non-covalent Interactions & UMA-Small & 0.84 & 2192/2206 \\
External & 2 & Non-covalent Interactions & MACE-OFF & 0.70 & 1728/2206 \\
External & 3 & Non-covalent Interactions & MACE-MP & 0.44 & 2206/2206 \\
\hline
Internal & 1 & Non-covalent Interactions & MACE-SPICE2 & 0.75 & 1807/2206 \\
Internal & 2 & Non-covalent Interactions & Visnet-SPICE2 & 0.73 & 1807/2206 \\
Internal & 2 & Non-covalent Interactions & Nequip-SPICE2 & 0.73 & 1807/2206 \\
Internal & 3 & Non-covalent Interactions & Visnet-SPICE2-t1x & 0.68 & 1807/2206 \\
Internal & 4 & Non-covalent Interactions & Nequip-t1x & 0.44 & 689/2206 \\
Internal & 5 & Non-covalent Interactions & MACE-t1x & 0.43 & 689/2206 \\
Internal & 6 & Non-covalent Interactions & Visnet-t1x & 0.21 & 689/2206 \\
\hline
External & 1 & Reactivity TST & UMA-Small &  0.86 & 11961/11961 \\
External & 1 & Reactivity TST & MACE-OFF &  0.12 & 11961/11961 \\
External & 1 & Reactivity TST & MACE-MP & 0.04 & 11961/11961 \\
\hline
Internal & 1 & Reactivity TST & Visnet-SPICE2-t1x  & 0.77 & 11961/11961 \\
Internal & 2 & Reactivity TST & Nequip-t1x & 0.44 & 11961/11961 \\
Internal & 3 & Reactivity TST & MACE-t1x & 0.43 & 11961/11961 \\
Internal & 4 & Reactivity TST & Visnet-t1x & 0.22 & 11961/11961 \\
Internal & 5 & Reactivity TST & MACE-SPICE2 & 0.10 & 11961/11961 \\
Internal & 6 & Reactivity TST & Visnet-SPICE2 & 0.05 & 11961/11961 \\
Internal & 7 & Reactivity TST & Nequip-SPICE2 & 0.04 & 11961/11961 \\
\hline
Internal & 1 & Nudged Elastic Band & Visnet-SPICE2-t1x & 0.58 & 100/100 \\
Internal & 1 & Nudged Elastic Band & Nequip-t1x & 0.58 & 100/100 \\
Internal & 2 & Nudged Elastic Band & MACE-t1x & 0.44 & 100/100 \\
Internal & 3 & Nudged Elastic Band & Visnet-t1x & 0.38 & 100/100 \\
\hline
\end{tabular}
\end{table}
\newpage
\begin{table}[h!]
\centering
\caption{Single benchmarks rankings (cont.)}
\label{tab:single-rankings-cont-2}
\begin{tabular}{c|c|c|c|c|c}
\hline
\textbf{Source} & \textbf{Rank} & \textbf{Benchmark} & \textbf{Model Name} & \textbf{Score} & \textbf{Test Cases} \\
\hline
External & 1 & Tautomers & UMA-Small &  0.23 & 1391/1391 \\
External & 2 & Tautomers & MACE-OFF &  0.07 & 1391/1391 \\
External & - & Tautomers & MACE-MP &  0.00 & 1391/1391 \\
\hline
Internal & 1 & Tautomers & Nequip-SPICE2 &  0.11 & 1391/1391 \\
Internal & 2 & Tautomers & Visnet-SPICE2 &  0.10 & 1391/1391 \\
Internal & 3 & Tautomers & Visnet-SPICE2-t1x &  0.09 & 1391/1391 \\
Internal & 3 & Tautomers & MACE-SPICE2 &  0.05 & 1391/1391 \\
\hline
External & 1 & Bond Length & UMA-Small & 1.00 & 8/8 \\
External & 1 & Bond Length & MACE-OFF & 1.00 & 8/8 \\
External & 1 & Bond Length & MACE-MP & 1.00 & 8/8 \\
\hline
Internal & 1 & Bond Length & Visnet-SPICE2-t1x & 1.00 & 8/8 \\
Internal & 1 & Bond Length & Visnet-SPICE2 & 1.00 & 8/8 \\
Internal & 1 & Bond Length & MACE-SPICE2 & 1.00 & 8/8 \\
Internal & 1 & Bond Length & Nequip-SPICE2 & 1.00 & 8/8 \\
\hline
External & 1 & Ring Planarity & MACE-OFF & 0.99 & 6/6 \\
External & 2 & Ring Planarity & UMA-Small & 0.98 & 6/6 \\
External & 1 & Ring Planarity & MACE-MP & 0.80 & 6/6 \\
\hline
Internal & 1 & Ring Planarity & Visnet-SPICE2-t1x & 1.00 & 6/6 \\
Internal & 1 & Ring Planarity & Visnet-SPICE2 & 1.00 & 6/6 \\
Internal & 1 & Ring Planarity & MACE-SPICE2 & 1.00 & 6/6 \\
Internal & 1 & Ring Planarity & Nequip-SPICE2 & 1.00 & 6/6 \\
\hline
\end{tabular}
\end{table}

\begin{figure}
    \centering
    \includegraphics[width=0.95\linewidth]{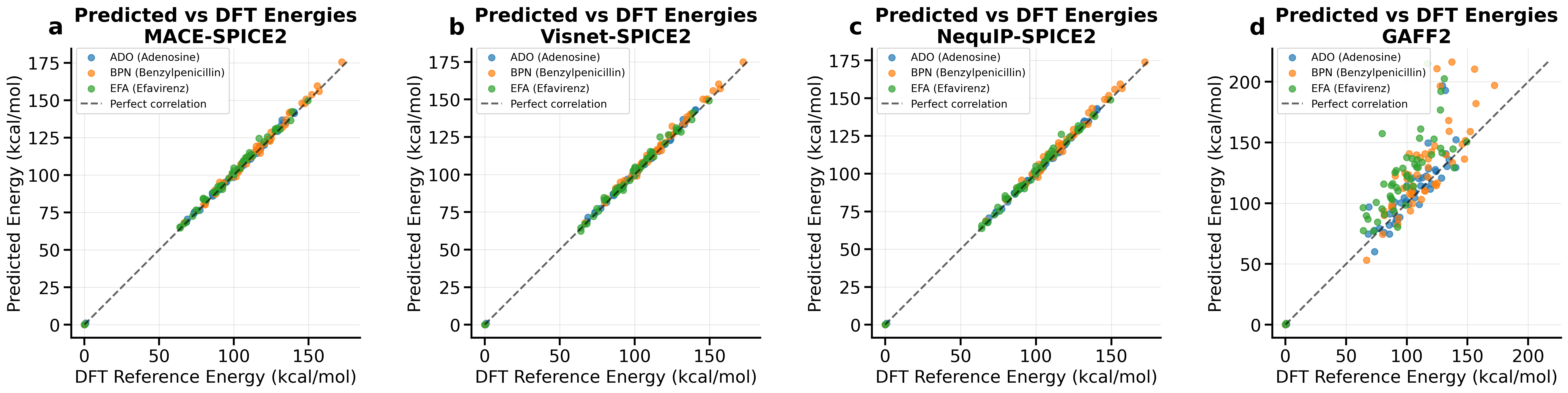}
    \caption{Predicted vs. DFT conformer energies for adenosine (ADO, blue), benzylpenicillin (BPN, orange), and efavirenz (EFA, green).}
    \label{fig:conformer-energy}
\end{figure}

\section{Road map for development}
We aim to release several updates to the current version (v0.1.0) in the coming months. These will include:

\begin{itemize}
    \item Additional metrics for molecular liquids.
    \item Material benchmarks.
    \item Additional MLIP models with a priority to models that have received validation through additional research.
    \item User contributed leaderboard.
\end{itemize}

Our objective is for any addition to the library to remain open source.

\end{document}